\documentclass[twocolumn,showpacs,superscriptaddress,aps,prb]{revtex4-1}
\usepackage{bm}
\usepackage{amssymb,amsmath}

\usepackage{graphicx}

\begin{document}

\title{Quantum critical behavior of the superfluid-Mott glass transition}

\author{Thomas Vojta}
\affiliation{Department of Physics, Missouri University of Science and Technology, Rolla, Missouri 65409, USA}

\author{Jack Crewse}
\affiliation{Department of Physics, Missouri University of Science and Technology, Rolla, Missouri 65409, USA}

\author{Martin Puschmann}
\affiliation{Department of Physics, Missouri University of Science and Technology, Rolla, Missouri 65409, USA}
\affiliation{Institute of Physics, Technische Universit\"at Chemnitz, 09107 Chemnitz, Germany}

\author{Daniel Arovas}
\affiliation{Department of Physics, University of California, San Diego, La Jolla, California 92093, USA}

\author{Yury Kiselev}
\affiliation{Department of Physics, University of California, San Diego, La Jolla, California 92093, USA}

\begin{abstract}
We investigate the zero-temperature superfluid to insulator transitions in a diluted two-dimensional quantum
rotor model with particle-hole symmetry. We map the Hamiltonian onto a classical $(2+1)$-dimensional
XY model with columnar disorder which we analyze by means of large-scale Monte Carlo simulations.
For dilutions below the lattice percolation threshold, the system undergoes a generic superfluid-Mott glass transition.
In contrast to other quantum phase transitions in disordered systems, its critical behavior is of
conventional power-law type with universal
(dilution-independent) critical exponents $z=1.52(3)$, $\nu=1.16(5)$, $\beta/\nu= 0.48(2)$, $\gamma/\nu=2.52(4)$,
and $\eta=-0.52(4)$.
These values agree with and improve upon earlier Monte-Carlo results [Phys. Rev. Lett. 92, 015703
(2004)] while (partially) excluding other findings in the literature.
As a further test of universality, we also consider a soft-spin version of the classical Hamiltonian.
In addition, we study the percolation quantum phase transition across the lattice percolation threshold;
its critical behavior is governed by the lattice percolation exponents in agreement with recent theoretical
predictions. We relate our results to a general classification of phase transitions in disordered systems,
and we briefly discuss experiments.
\end{abstract}

\date{\today}

\maketitle
\section{Introduction}
\label{Intro}

Zero-temperature phase transitions between superfluid and insulating ground states
in systems of disordered interacting bosons are prototypical quantum phase transitions with
experimental applications ranging from helium absorbed in vycor \cite{CHSTR83,Reppy84} to Josephson
junction arrays \cite{ZFEM92,ZEGM96}, superconducting films
\cite{HavilandLiuGoldman89,HebardPaalanen90}, doped quantum magnets in high fields
\cite{OosawaTanaka02,HZMR10,Yuetal12}, and to ultracold atoms in disordered optical
lattices \cite{WPMZCD09,KSMBE13,DTGRMGIM14}.

For generic disorder, the two bulk phases, viz.\ superfluid and Mott insulator,
are separated by another phase, the Bose glass which is a compressible gapless
insulator \cite{FisherFisher88,FWGF89,PPST09}. It can be understood as the Griffiths phase
\cite{Griffiths69,ThillHuse95,Vojta06}
of the superfluid-insulator transition in which rare large regions of
local superfluid order coexist with the insulating bulk.
The quantum phase transition between superfluid and Bose
glass has been studied in great detail using various analytical and computational
techniques. It has recently reattracted considerable attention because new analytical
\cite{WeichmanMukhopadhyay07} and numerical \cite{PCLB06,MeierWallin12,NgSorensen15,ALLL15}
findings have challenged the scaling relation\cite{FisherFisher88,FWGF89}  $z=d$
between the dynamical exponent $z$ and the space dimensionality $d$
(Refs.\ \onlinecite{WeichmanMukhopadhyay07,PCLB06,MeierWallin12,NgSorensen15,ALLL15}
also contain long lists of references to earlier work.)

In the presence of particle-hole symmetry, the glassy Griffiths phase between superfluid and
Mott insulator has a different character: it is the incompressible gapless Mott glass
(also called the random-rod glass) \cite{GiamarchiLeDoussalOrignac01,WeichmanMukhopadhyay08}.
The quantum phase transition between superfluid and Mott glass has attracted less
attention than the Bose glass transition. Moreover, the available quantitative results
for two space dimensions do not agree with each other. Monte Carlo simulations
of a link-current model \cite{ProkofevSvistunov04} yielded a dynamical critical
exponent $z=1.5(2)$ and a correlation function exponent $\eta=-0.3(1)$.
\footnote{Here, the numbers in parentheses are the errors of the last digits.} A numerical strong-disorder
renormalization group study of a particle-hole symmetric quantum rotor model gave  $z=1.31(7)$,
a correlation length exponent $\nu=1.09(4)$,
and $\gamma/\nu= 1.1(2)$ where $\gamma$ is the order parameter susceptibility exponent
\cite{IyerPekkerRefael12}.
The Fisher relation $2-\eta=\gamma/\nu$ then implies $\eta=0.9(2)$. Furthermore, a recent
Monte Carlo study of a quantum rotor model\cite{SLRT14} reported good scaling by setting $z$ to
its clean value $z=1$ which resulted
in $\nu=0.96(6)$.
All these models are expected to be in the same universality class.
The critical behavior of the superfluid-Mott glass
quantum phase transition in two dimensions must thus be considered an open question.

To address this question, we consider a site-diluted two-dimensional quantum
rotor model with particle-hole symmetry. After mapping this Hamiltonian onto
a classical $(2+1)$-dimensional XY model with columnar defects, we perform large-scale
Monte Carlo simulations for lattices with up to 11 million sites, averaging
over $10\,000$ to $50\,000$ disorder configurations. The data are analyzed by a finite-size
scaling technique\cite{GuoBhattHuse94,RiegerYoung94,SknepnekVojtaVojta04,*VojtaSknepnek06}
that does not require prior knowledge of the dynamical exponent $z$. We also
include the leading corrections to scaling. Our results can be summarized as follows:
The system features two distinct quantum phase transitions.
For dilutions $p$ below the percolation threshold $p_c$ of the lattice, we find a
superfluid-Mott glass transition characterized by universal (dilution-independent)
critical behavior with exponent values $z=1.52(3)$, $\nu=1.16(5)$, $\beta/\nu= 0.48(2)$,
$\gamma/\nu=2.52(4)$, and $\eta=-0.52(4)$. The transition across the lattice percolation threshold $p_c$
falls into a different universality class. Its simulation data can be fitted well
with the theory developed in Ref.\ \onlinecite{VojtaSchmalian05b} which yields critical exponents
that can be expressed in terms of the classical percolation exponents and take the rational values
$z=91/48$, $\beta/\nu= 5/48$, $\gamma/\nu=59/16$, and $\eta=-27/16$.

The rest of the paper is organized as follows. Section \ref{Model} introduces
the quantum rotor Hamiltonian, the mapping to the classical XY model, and the
finite-size scaling technique.
Monte Carlo simulations for both the generic ($p<p_c$) transition and the percolation
transition are discussed in Sec.\ \ref{sec:MC}.
We conclude in Sec.\ \ref{sec:Conclusions}.

\section{Theory}
\label{Model}
\subsection{Diluted rotor model}
\label{subsec:rotor}

The starting point is a site-diluted quantum rotor model on a square lattice given by the
Hamiltonian
\begin{equation}
 H = \frac U 2 \sum_i \epsilon_i (\hat n_i - \bar n_i)^2 -J\sum_{\langle ij \rangle} \epsilon_i \epsilon_j \cos(\hat \phi_{i}-\hat \phi_j)~.
\label{eq:Hamiltonian}
\end{equation}
Here, $\hat n_i$ is the number operator at site $i$, $\hat \phi_i$ is the phase operator,
and $U$ and $J$ represent the charging energy and the Josephson coupling, respectively.
$\bar n_i$ is the offset charge at site $i$. In the Josephson term,
$\langle ij \rangle$ refers to pairs of nearest neighbors. The quenched random variables $\epsilon_i$
implement the site dilution. They are independent of each other and
take the values 0 (vacancy) with probability $p$ and 1 (occupied site) with probability $1-p$.

As we are interested in the superfluid-Mott glass transition, we set all offset charges $\bar n_i$
to zero and consider commensurate (integer) filling $\langle \hat n \rangle$. In this case, the disorder is purely
off-diagonal, and the model is particle-hole symmetric. The qualitative features of its phase diagram
are well understood \cite{FWGF89,WeichmanMukhopadhyay08}. If the charging energy dominates, $U \gg J$,
the ground state is a Mott insulator. In the opposite limit, $J \gg U$,
the ground state is a superfluid as long as the dilution $p$ is below the lattice percolation threshold
$p_c$. For $p>p_c$, the lattice consists of disconnected clusters and a long-range ordered superfluid state
is impossible.

\begin{figure}
\centerline{
\includegraphics[width=0.65\columnwidth]{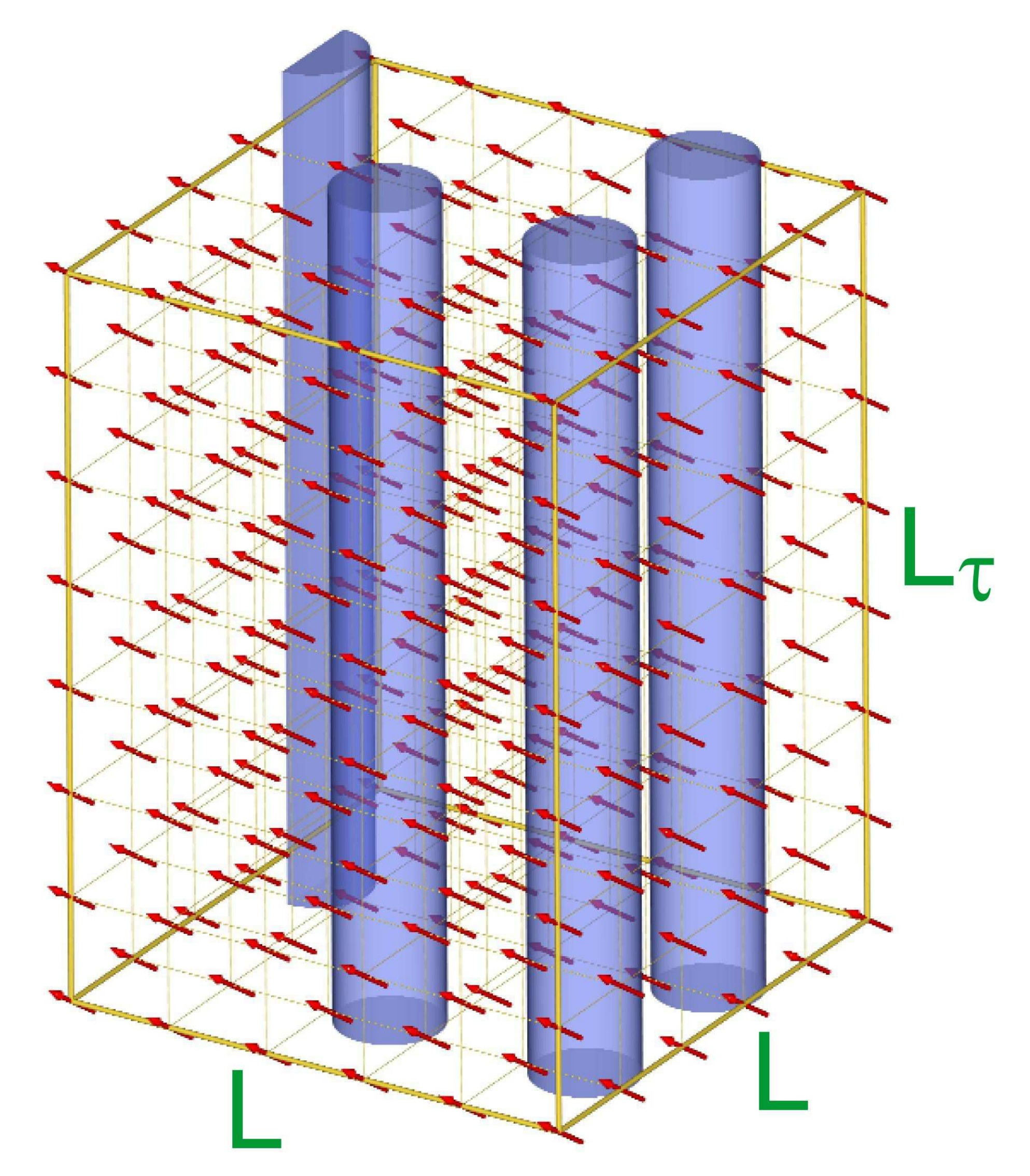}
} \caption{Sketch of the classical XY model (\ref{eq:Hcl}). The arrows represent the classical
unit vectors $\mathbf{S}$, and the tubes show the locations of the vacancy columns.}
\label{fig:swisscheese}
\end{figure}

In the case of particle-hole symmetry, the quantum rotor model (\ref{eq:Hamiltonian}) can be mapped \cite{WSGY94}
onto a classical $(2+1)$-dimensional XY model on a cubic lattice having the Hamiltonian
\begin{equation}
 H_{\rm cl} = -J_s\sum_{\langle i,j\rangle, t}
\epsilon_i\epsilon_j\mathbf{S}_{i,t}\cdot\mathbf{S}_{j,t}
-J_\tau \sum_{i,t}\epsilon_i\mathbf{S}_{i,t}\cdot\mathbf{S}_{i,t+1}
\label{eq:Hcl}
\end{equation}
where $\mathbf{S}_{i,t}$ is an O(2) unit vector at the lattice site with spatial
coordinate $i$ and ``imaginary time'' coordinate $t$.
The coupling constants $J_s/T$ and $J_\tau/T$ are determined by the original quantum rotor
Hamiltonian (\ref{eq:Hamiltonian}) with $T$ being an effective
``classical'' temperature, not equal to the real physical temperature. (The physical temperature of the quantum
system (\ref{eq:Hamiltonian}) maps onto the inverse system size in imaginary time direction of the classical model.)
Due to universality,
the exact values of $J_s$ and $J_\tau$ are not important for the critical behavior. We therefore set $J_s=J_\tau=1$
and drive the XY model (\ref{eq:Hcl}) through the transition by varying the classical temperature $T$.
Because the vacancy positions do not depend on the imaginary time coordinate $t$, the
defects in the classical model (\ref{eq:Hcl}) are columnar, i.e., the disorder is perfectly correlated in
the imaginary time direction (see Fig.\ \ref{fig:swisscheese}).

In the clean undiluted limit $p=0$, the Hamiltonian (\ref{eq:Hcl}) simplifies to the usual three-dimensional
XY model. The correlation length critical exponent of the three-dimensional XY universality class takes the
value $\nu\approx 0.6717$ (see, e.g., Ref.\ \onlinecite{CHPV06}). This value violates the Harris criterion
\cite{Harris74} $d \nu > 2$ where $d=2$ is the number of dimensions in which there is randomness.
Consequently, the three-dimensional clean XY critical point is unstable against columnar defects, and we
expect the diluted system to feature a different critical behavior.

\subsection{Anisotropic finite-size scaling}
\label{subsec:FSS}

Finite-size scaling \cite{Barber_review83,Cardy_book88} is a powerful tool for analyzing
Monte Carlo data. Particularly useful are quantities of scale dimension zero such as the
(average) Binder cumulant
\begin{equation}
\label{eq:Binder_av} g_{\rm av}=\left[ 1-\frac{\langle |\mathbf{m}|^4\rangle}{3\langle
|\mathbf{m}|^2\rangle^2}\right]_{\rm dis},
\end{equation}
where $\mathbf{m} = (1/N)\sum_{i,\tau}\mathbf{S}_{i,\tau}$ is the order parameter
($N$ denotes the number of lattice sites).
 $\left[\ldots\right]_{\rm dis}$ refers to the disorder average and $\langle\ldots\rangle$
denotes the Monte Carlo average for each sample. In an isotropic system with a single
relevant length scale, it takes the scaling form $g_{\rm av}(r,L) =  X(rL^{1/\nu})$.
Here $L$ is the linear system size, $r=(T-T_c)/T_c$ is the distance from criticality, and
$X$ is a scaling function. This scaling form implies that $g_{\rm av}$ vs.\ $r$
curves for systems of different sizes $L$ all cross at criticality, $r=0$,
having the value $g_{\rm av}(0,L) = X(0)$.
This can be used to find the critical point with high accuracy. Moreover, the slopes
of the $g_{\rm av}$ vs.\ $r$ curves at $r=0$ vary as $L^{1/\nu}$ which can be used to measure
$\nu$.

As the quenched disorder in our Hamiltonian (\ref{eq:Hcl}) breaks the symmetry between the
space and imaginary time directions, we need to distinguish the linear system size
$L$ in the two space directions from the size $L_\tau$ in the imaginary time direction.
($L_\tau$ corresponds to the inverse physical temperature of the original quantum model
(\ref{eq:Hamiltonian}).)
If the putative disordered critical point fulfills conventional power-law dynamical scaling,
the finite-size scaling form of the average Binder cumulant then reads
\begin{equation}
g_{\rm av}(r,L,L_\tau) = X_{g_{\rm av}}(rL^{1/\nu},L_\tau/L^z)
\label{eq:Binder_FSS}
\end{equation}
where $z$ is the dynamical critical exponent, and $X_{g_{\rm av}}$ is the dimensionless scaling function
which now depends on two arguments.
Note that some quantum phase transitions
in disordered systems feature exotic activated dynamical scaling instead of
power-law scaling, for example the ferromagnetic transition in the random transverse-field Ising
model \cite{Fisher92,*Fisher95}, the pairbreaking superconductor-metal quantum phase transition
\cite{HoyosKotabageVojta07,*VojtaKotabageHoyos09,DRMS08,*DRHV10,Xingetal15}, and magnetic transitions
in itinerant systems \cite{UbaidKassisVojtaSchroeder10,Vojta10}. For activated dynamical scaling,
the scaling combination $L_\tau/L^z$ in Eq.\ (\ref{eq:Binder_FSS}) needs to be replaced
by $\ln(L_\tau)/L^\psi$ where $\psi$ is the tunneling exponent. Based on the classification
of disordered quantum phase transitions developed in Refs.\ \onlinecite{VojtaSchmalian05,Vojta06},
we do not expect the superfluid-Mott glass transition to show activated scaling. We will return
to this point in the concluding section.

How can one perform a finite-size scaling analysis of Monte Carlo data based on the
scaling form (\ref{eq:Binder_FSS}) of the average Binder cumulant? If the value of $z$
is known, the analysis is as simple as in the isotropic case: One chooses system sizes $L$
and $L_\tau$ such that $L_\tau =c\, L^z$ were $c$ is a constant. Then
the  $g_{\rm av}$ vs.\ $r$ curves for systems of different sizes cross at criticality
[with the value $g_{\rm av}(0,L,c\,L^z) = X_{g_{\rm av}}(0,c)$] which can be used to locate the critical point.
However, in the absence of a value for $z$, this approach breaks down because the correct
shapes (aspect ratios) of the samples are not known.

A method for finding the correct sample shape within the simulations
\cite{GuoBhattHuse94,RiegerYoung94,SknepnekVojtaVojta04,*VojtaSknepnek06} can be based on the
following property of the Binder cumulant: For fixed $L$, $g_{\rm av}$ as a
function of $L_{\tau}$ has a peak at position $L_{\tau}^{\rm max}$ and value $g_{\rm av}^{\rm max}$.
The peak position marks the {\em optimal}
sample shape, where the ratio $L_{\tau}/L$ behaves like the corresponding ratio
of the correlation lengths in time and space directions, $\xi_{\tau}/\xi_s$. (If the aspect
ratio deviates from the optimal one, the system can be decomposed into independent units
either in space or in time direction, and thus $g_{\rm av}$ decreases.) At criticality,
$L_{\tau}^{\rm max}$ must be proportional to $L^z$, fixing the second
argument of the scaling function $X_{g_{\rm av}}$. This implies that the peak value $g_{\rm av}^{\rm max}$ at criticality
is independent of $L$ and that the  $g_{\rm av}$ vs.\ $r$ curves of samples of the optimal shape
($L_\tau=L_\tau^{\rm max}$) cross at $r=0$.

In our simulations, we use an iterative approach. We start from a guess for $z$ and the corresponding
sample shapes. The approximate crossing of the  $g_{\rm av}$ vs.\ $r$ curves for these samples gives
an estimate for $T_c$. At this temperature, we next analyze $g_{\rm av}$ as a function of $L_\tau$
for fixed $L$. The values of $L_\tau^{\rm max}$ give improved estimates for the optimal sample shapes
and thus for $z$. After iterating this
procedure three or four times, the values of $T_c$ and $z$ will have converged with reasonable
accuracy.

Once $z$ and $T_c$ are determined, the finite-size scaling analysis proceeds as usual, based
on the scaling forms
\begin{eqnarray}
m &=& L^{-\beta/\nu} X_m(rL^{1/\nu},L_\tau/L^z)~, \label{eq:m_FSS}\\
\chi &=& L^{\gamma/\nu} X_\chi(rL^{1/\nu},L_\tau/L^z) \label{eq:chi_FSS}
\end{eqnarray}
for the order parameter $m$ and its susceptibility $\chi$. Here, $X_m$ and $X_\chi$
are dimensionless scaling functions, and $\beta$ and $\gamma$ are the order parameter
and susceptibility critical exponents, respectively.

In addition to these thermodynamic
quantities, we also calculate the correlation lengths $\xi_s$ and $\xi_\tau$ is the space
and imaginary time directions, respectively. They are obtained, as usual, from the second moment
of the spin-spin correlation function \cite{CooperFreedmanPreston82,Kim93,CGGP01}
and can be expressed in terms of the Fourier transform $\tilde G(q_s,q_\tau)$ of the correlation function,
\begin{eqnarray}
\xi_s &=& \left[ \left(\frac{\tilde G(0,0) -\tilde G (q_{s0},0)}{q_{s0}^2 \, \tilde G(q_{s0},0)} \right)^{1/2} \right]_\textrm{dis}~,
\label{eq:xis}\\
\xi_\tau &=& \left[ \left(\frac{\tilde G(0,0) -\tilde G (0,q_{\tau0})}{q_{\tau0}^2 \, \tilde G(0,q_{\tau0})} \right)^{1/2} \right]_\textrm{dis}~.
\label{eq:xit}
\end{eqnarray}
Here, $q_{s0}=2\pi/L$ and $q_{\tau0}=2\pi/L_\tau$ are the minimum values of the wave numbers
$q_s$ and $q_\tau$ that fit into a system of linear size $L$ and $L_\tau$ in space and imaginary
time direction, respectively. The reduced correlation lengths $\xi_s/L$ and $\xi_\tau/L_\tau$ have scale dimension zero,
their scaling forms therefore read
\begin{eqnarray}
\xi_s/L &=& X_{\xi_s}(rL^{1/\nu},L_\tau/L^z)~, \label{eq:xis_FSS}\\
\xi_\tau/L_\tau &=&  X_{\xi_\tau}(rL^{1/\nu},L_\tau/L^z)~. \label{eq:xit_FSS}
\end{eqnarray}

\section{Monte Carlo simulations}
\label{sec:MC}
\subsection{Overview}
\label{subsec:MC-Overview}

Our Monte Carlo simulations of the classical XY model (\ref{eq:Hcl}) combine the Wolff
cluster algorithm \cite{Wolff89} with conventional Metropolis updates \cite{MRRT53}.
Specifically, a full Monte Carlo sweep consists of a Metropolis sweep over the lattice
followed by a Wolff sweep. (A Wolff sweep is defined as a number of
cluster flips such that the total number of flipped spins equals the number of
lattice sites.) The Wolff algorithm greatly reduces the critical slowing down, and the
Metropolis updates equilibrate small disconnected clusters of sites that are missed
in the construction of the Wolff clusters (this becomes important at higher dilutions $p$).

We simulate systems with linear sizes up to $L=150$ in space direction and
up to $L_\tau=1792$ in the imaginary time direction at dilutions
$p=0$, 1/8, 1/5, 2/7, 1/3, 9/25 and the percolation threshold $p_c=0.407253$.

The simulation of disordered systems requires a high numerical effort because
many samples with different disorder configurations need to be studied to compute
averages, variances, and distributions of observables. For good performance,
one must thus carefully optimize the number $n_s$ of samples (i.e., disorder configurations) and the number
$n_m$ of measurements during the simulation of each sample.
Based on the consideration in Refs.\ \onlinecite{BFMM98,BFMM98b,SknepnekVojtaVojta04,*VojtaSknepnek06,ZWNHV15},
we have chosen rather short runs of $n_m=500$ full sweeps per sample (with a measurement after each sweep)
but large numbers of disorder configurations ranging from $n_s=10\,000$ to 50\,000 depending on the system size.
The equilibration period is taken to be 100 full sweeps, significantly longer than the actual
equilibration times that reach 30 to 40 sweeps at maximum.
Short Monte Carlo runs can lead to biases in some of the observables.
To eliminate these, we have implemented improved estimators along the lines
discussed in the appendix of Ref.\ \onlinecite{ZWNHV15}.

\begin{figure}
\includegraphics[width=8.6cm]{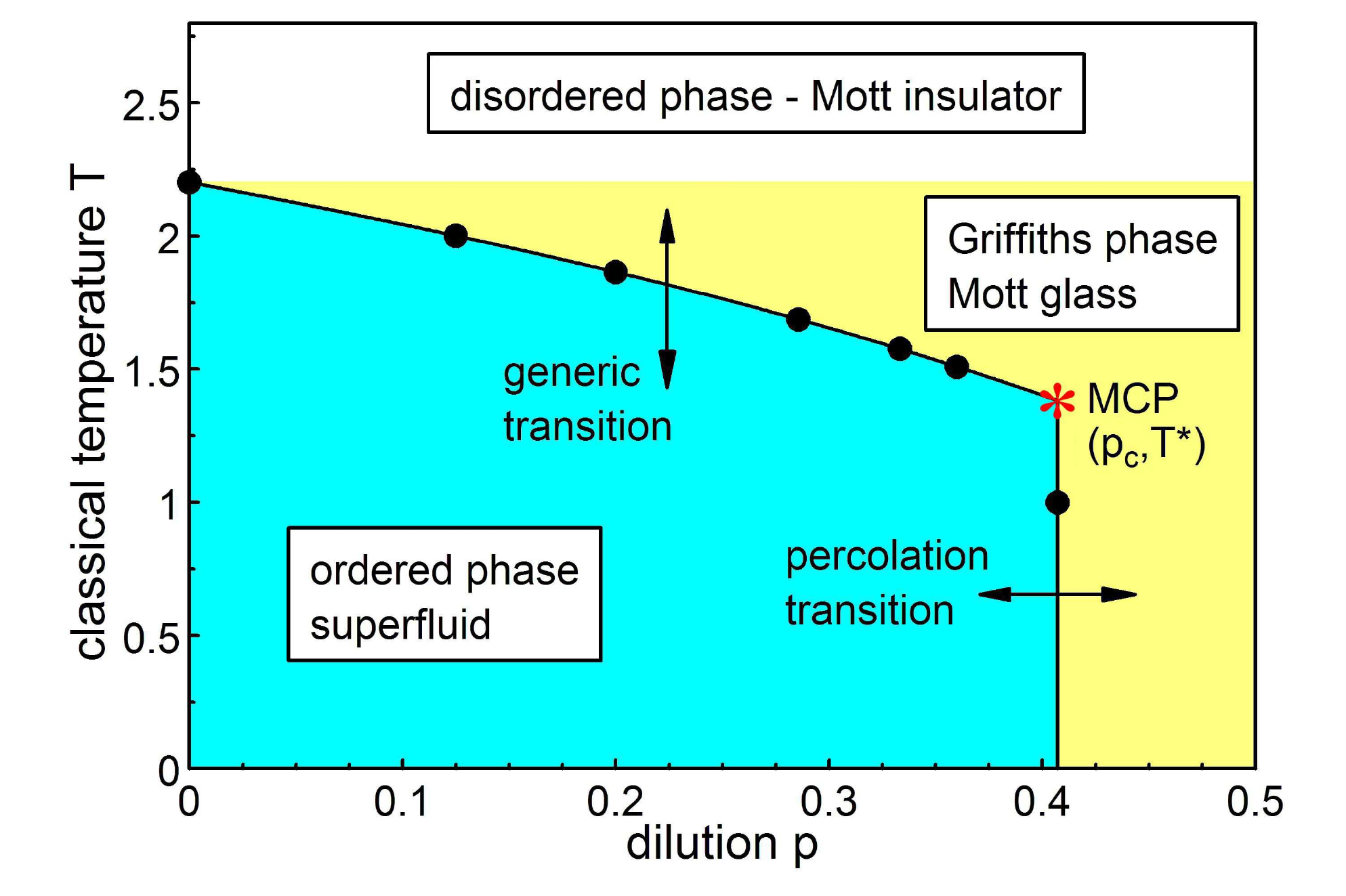}
\caption{(Color online) Phase diagram of the classical XY model (\ref{eq:Hcl})
as a function of classical temperature and dilution. MCP is the multicritical point
that separates the generic and percolation transitions.
The big dots mark the numerically determined transition points. The lines
are guides for the eye only.}
\label{fig:pd}
\end{figure}

The phase diagram resulting from these simulations is shown in Fig.\ \ref{fig:pd}.
The critical temperature $T_c(p)$ decreases with increasing dilution from its clean
value $T_c(0)$, as expected. For dilutions above the percolation threshold
$p_c=0.407253$, the lattice consists of disconnected finite-size clusters.
Therefore, long-range superfluid order is impossible. Right at $p_c$, there is an
infinite cluster of dimension $1+d_f$ where $d_f=91/48$ is the dimensionality of the
critical percolation cluster in two dimensions, and the extra 1 stems from the
imaginary time direction. As $1+d_f$ is larger than the lower critical dimension
$d_c^-=2$ of the XY model, the XY model on the critical percolation cluster orders
below a multicritical temperature $T^\ast$. This implies that the phase boundary
coincides with the classical percolation threshold for $T<T^\ast$ (see also Ref.\ \onlinecite{VojtaHoyos08b}).
We thus identify two different phase transitions,
(i) the generic superfluid-Mott glass transition for $p<p_c$ and (ii) a percolation transition
across the lattice percolation threshold.

In the following sections, we discuss the critical behaviors of these transitions
in detail. To test our codes, we have also studied the clean limit $p=0$
using system sizes up to $224^3$ sites. By analyzing the crossings
of the Binder cumulant and the reduced correlation length, we find a critical temperature
$T_c(0)=2.201844(4)$. Finite-size scaling then gives the critical exponents
$\beta/\nu=0.518(3)$, $\gamma/\nu=1.961(3)$, and $\nu=0.673(2)$. Within their errors,
they agree well with high-precision results for the three-dimensional XY
universality class \cite{CHPV06}.

As a further test for the universality of the (generic) critical behavior,
we also perform exploratory simulations of a soft-spin version of the
classical Hamiltonian. They are discussed in Sec.\ \ref{subsec:MC-soft}.

\subsection{Generic superfluid-Mott glass transition}
\label{subsec:MC-Generic}

To analyze the critical behavior of the generic transition occurring
for $0<p<p_c$, we consider five different dilutions,
$p=1/8$, 1/5, 2/7, 1/3, and 9/25. As described in Sec.\ \ref{subsec:FSS}, we use
an iterative procedure that consists of two types of simulation runs. The first
are runs right at $T_c$ for systems with several different $L_\tau$ for each $L$.
Finite-size scaling of the Binder cumulant at $T_c$
as a function of $L$ and $L_\tau$ gives the optimal sample shapes and the dynamical
exponent $z$.  In the second set of simulations, we vary the temperature over a
range in the vicinity of $T_c$, but we consider only the optimal shapes found
in the first part. Finite-size scaling of the order parameter, susceptibility, Binder
cumulant, and correlation length as functions of $L$ and $T$ then yields the
critical exponents $\beta/\nu$, $\gamma/\nu$, and $\nu$.

\begin{figure}
\includegraphics[width=8.6cm]{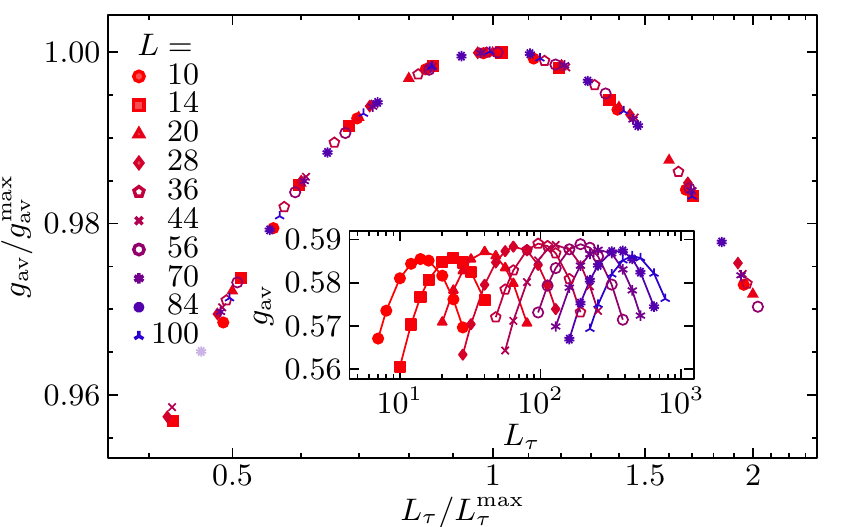}
\caption{(Color online) Binder cumulant $g_{\rm av}$ as a function of $L_\tau$ for several $L$
        at the critical temperature $T_c=1.577$ for dilution $p=1/3$. The relative statistical error of
        $g_{\rm av}$ is between 0.05\% and 0.1\%. Inset: Raw data $g_{\rm av}$ vs.\ $L_\tau$.
        Main panel: Scaling plot $g_{\rm av}/g_{\rm av}^{\rm max}$
        vs.\ $L_\tau/L_\tau^{\rm max}$.}
\label{fig:g_L_Lt}
\end{figure}

The inset of Fig.\ \ref{fig:g_L_Lt} shows the Binder cumulant $g_{\rm av}$ as a function of $L_\tau$
for several $L=10$ to 100 at the estimated critical temperature $T_c=1.577$ for dilution
$p=1/3$.
As expected at the critical point, the maximum Binder cumulant $g_{\rm av}^{\rm max}$ for each of the curves
does not depend on $L$. (The remaining weak variation visible in the figure can be attributed
to corrections to scaling, see below.)
To generate a scaling plot that tests the scaling form (\ref{eq:Binder_FSS}), we now fit each
$g_{\rm av}$ vs.\ $L_\tau$ curve with an inverted parabola in $\ln L_\tau$. The vertex of this parabola
yields the position $L_\tau^{\rm max}$ of the maximum and its value $g_{\rm av}^{\rm max}$.
When plotting  $g_{\rm av}/g_{\rm av}^{\rm max}$ vs.\ $L_\tau/L_\tau^{\rm max}$ the data scale very well,
as can be seen in the resulting scaling plot in the main panel of Fig.\ \ref{fig:g_L_Lt}. This
demonstrates that the Binder cumulant fulfills Eq.\ (\ref{eq:Binder_FSS})
with high accuracy. We have created the corresponding scaling plots for all the other dilutions,
$p=1/8$, 1/5, 2/7, and 9/25, with analogous results.
\footnote{For low dilutions $p$, the parabola fits of $g_{\rm av}$ vs.\ $L_\tau$ 
are affected by corrections to scaling for 
small $L$ and $L_\tau$. We thus slightly adjust $L_\tau^{\rm max}$ and $g_{\rm av}^{\rm max}$ 
to further improve the quality of the data collapse onto a common master curve.
This applies to the four smallest system sizes $L$ for $p=1/8$ and 1/5 and the three smallest sizes
for $p=2/7$. The resulting change of the value of $z$ is about 0.01, well below the
error due to the uncertainty in $T_c$. }

\begin{figure}
\includegraphics[width=8.6cm]{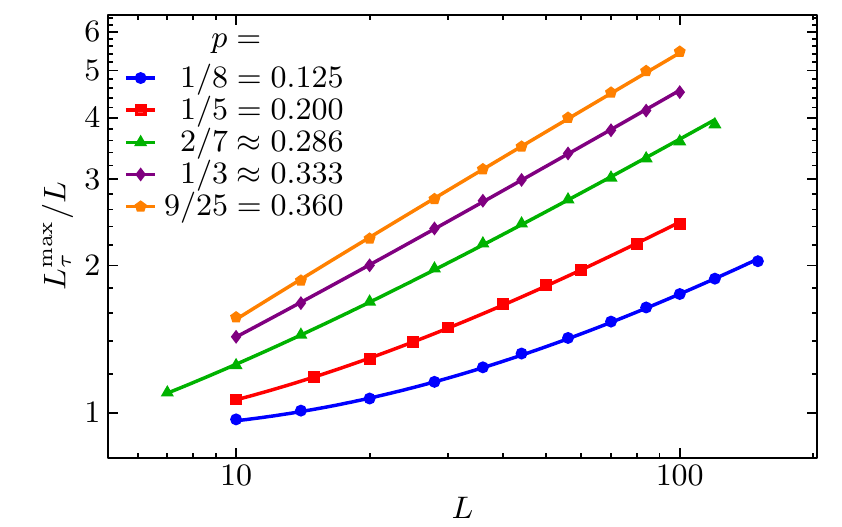}
\caption{(Color online) Double logarithmic plot of $L_\tau^{\rm max}/L$ vs.\ $L$ for several dilutions $p$ below
        the percolation threshold. Solid lines at fits to $L_\tau^{\rm max}=aL^z(1+bL^{-\omega})$ giving
$z=1.526(5)$ and $\omega=0.76(2)$. The statistical errors of the data are well below a
symbol size (The statistical error of $L_\tau^{\rm max}$ is determined by repeating the scaling analysis
for 1000 synthetic data sets that add to the original data set a Gaussian random noise that corresponds to
the uncertainties of the data.) }
\label{fig:scaling_z}
\end{figure}

To determine the dynamical critical exponent $z$, we now analyze the dependence
of the positions $L_\tau^{\rm max}$ of the maximum on $L$.
According to Eq.\ (\ref{eq:Binder_FSS}),
we expect the power-law dependence $L_\tau^{\rm max} \sim L^z$.
In Fig.\ \ref{fig:scaling_z}, we plot $L_\tau^{\rm max}$ vs.\ $L$ for all dilutions $p<p_c$.
The curves show significant deviations from pure power-law behavior, in particular for the smaller dilutions,
indicating that the crossover from clean to disordered critical behavior is slow. The resulting
corrections to scaling are strong and cannot be neglected. Pure power-law fits of the data
would therefore only yield effective, scale-dependent exponents. To determine the true asymptotic exponents,
we include the leading corrections to scaling via the ansatz $L_\tau^{\rm max}=aL^z(1+bL^{-\omega})$
with universal (dilution-independent) critical exponents $z$ and $\omega$ but dilution-dependent
prefactors $a$ and $b$. The exponent values resulting from a combined fit of the data for all five
dilutions are $z=1.526(5)$ and $\omega=0.76(2)$. The fit is of good quality giving $\tilde \chi^2\approx 1.4$.
[We denote the reduced sum of squared errors of the fit (per degree of freedom)
by $\tilde \chi^2$ to distinguish it from the susceptibility $\chi$.]
The fit is also robust against removing complete data sets or removing points from the upper or lower end of each set.
Interestingly, the leading corrections to scaling appear to vanish somewhere between $p=1/3$ and 9/25,
as the prefactor $b$ of the correction term changes sign. Correspondingly, pure power-law fits
of the  $p=1/3$ and 9/25 data yield $z=1.502$ and 1.546, respectively. These values are close to the estimate from
the combined fit and nicely bracket it on both sides. An additional significant source of errors is the
uncertainty of the critical
temperature. To assess its effect on the dynamical exponent, we repeat the $L_\tau^{\rm max}$ vs.\ $L$
analysis (for dilutions $p=1/3$ and 9/25) at temperatures slightly above and below our estimated $T_c$
($\Delta T_c\approx 0.003$, roughly at the boundaries of our confidence intervals). This leads to shifts in $z$ of about
0.01 to 0.02. Our final estimate for the dynamical critical exponent therefore reads
$z=1.52(3)$.

\begin{figure}
\includegraphics[width=8.6cm]{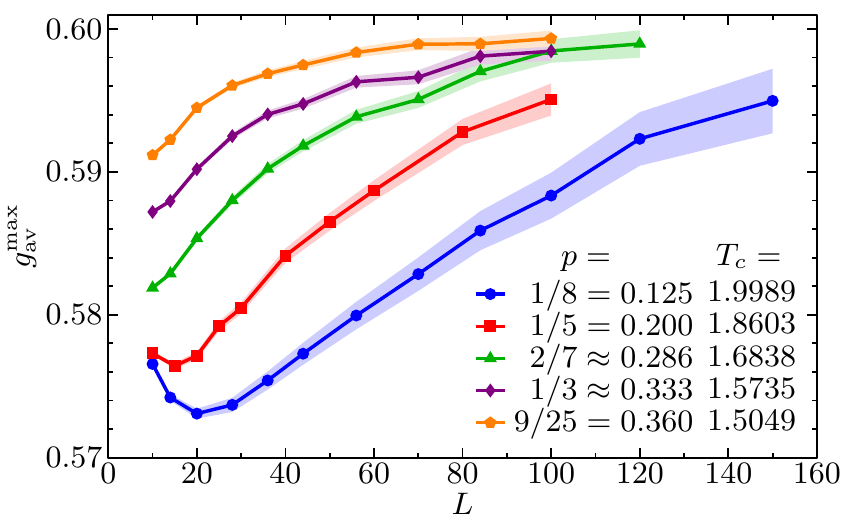}
\caption{(Color online)  $g_{\rm av}^{\rm max}$ vs.\ $L$ at the improved estimates for $T_c$.
The statistical errors of the data points are about a symbol size or smaller.
The shading represents the range of $g_{\rm av}^{\rm max}$ values for temperatures $T$ within $T_c\pm 0.0002$
and is intended to illustrate to what extent the extrapolation depends on $T$.
Based on these data we estimate that
the error of $T_c$ does not exceed 0.001. }
\label{fig:critT}
\end{figure}

To find the remaining critical exponents, we now turn to the Monte Carlo runs that use the optimal
sample shapes $(L,L_\tau^{\rm max})$. According to Eqs.\  (\ref{eq:m_FSS}) and (\ref{eq:chi_FSS}), $\beta/\nu$
and $\gamma/\nu$  can be
obtained from the $L$ dependence of the order parameter and susceptibility at $T_c$ of the optimally shaped
samples. As we expect corrections to scaling to be important, we again include
subleading terms in our fit functions,  $m=a L^{-\beta/\nu}(1+b L^{-\omega})$ for the order parameter and
$\chi=a L^{\gamma/\nu}(1+b L^{-\omega})$ for the susceptibility. Here $\beta/\nu$, $\gamma/\nu$, and $\omega$
are the universal, dilution-independent critical exponents while the coefficients $a$ and $b$ are again
non-universal. (Note that $a$ and $b$ generally differ from quantity to quantity; we use the same symbols
to avoid cluttering up the notation too much.)
When performing fits of our data to
these expressions, we noticed, however, that the quality of the fits is extremely sensitive to small
changes of the estimates for $T_c$ (much more so than in the analysis of the dynamical exponent $z$ above).
To determine higher accuracy estimates of $T_c$, we use the criterion that the value
of $g_{\rm av}^{\rm max}$ at criticality should approach a \emph{dilution-independent} constant with
$L\to \infty$ at a universal critical point. Varying $T$ until this criterion is fulfilled yields
improved estimates for the critical temperatures, viz.\ $T_c=1.9989$ for $p=1/8$,
$T_c=1.8603$ for $p=1/5$, $T_c=1.6838$ for $p=2/7$, $T_c=1.5735$ for $p=1/3$, and
$T_c=1.5049$ for $p=9/25$. We estimate the error of these values to be about $0.001$.
Figure \ref{fig:critT} shows the resulting dependence $g_{\rm av}^{\rm max}$ on $L$.
In the large-$L$ limit, $g_{\rm av}^{\rm max}$ approaches the value 0.599(2).
Note that the non-monotonic behavior of $g_{\rm av}^{\rm max}$ for weak
dilutions suggests that at least two corrections to scaling terms contribute
at small $L$.

\begin{figure}
\includegraphics[width=8.6cm]{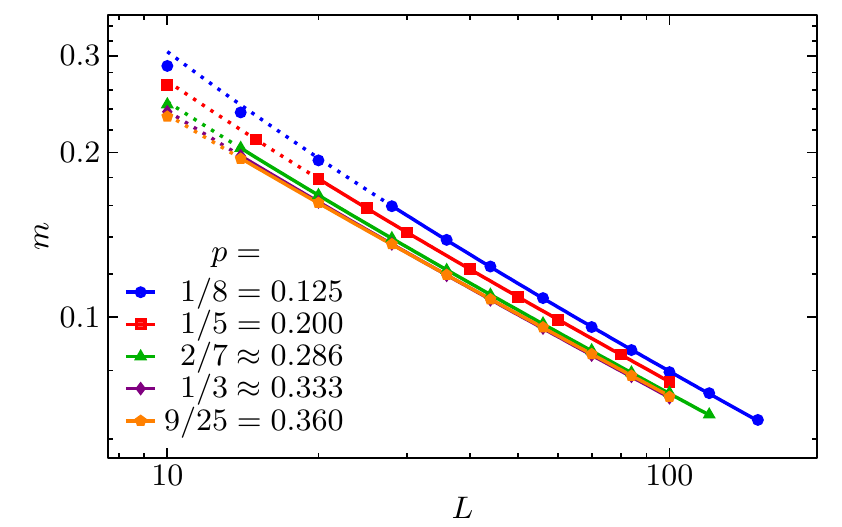}
\caption{(Color online) Double logarithmic plot of $m$ vs.\ $L$ for several dilutions $p$ below
        the percolation threshold. Solid lines at fits to $m=a L^{-\beta/\nu}(1+b L^{-\omega})$ giving
    $\beta/\nu=0.480(8)$ and $\omega=0.82(2)$. The lines are dotted in the regions not included
    in the fit. The statistical errors of the data are well below a
symbol size. }
\label{fig:scaling_beta}
\end{figure}

Using the improved critical temperatures, we now proceed to determine $\beta/\nu$
and $\gamma/\nu$.  Figure \ref{fig:scaling_beta} shows the order parameter $m$ at $T_c$
as a function of $L$ for all dilutions $p<p_c$.
The combined fit of all data to $m=a L^{-\beta/\nu}(1+b L^{-\omega})$ is of good quality
($\tilde \chi^2\approx 0.64$) if the smallest system sizes are excluded (see figure).
Interestingly, the sizes that need to be excluded
are exactly those for which  $g_{\rm av}^{\rm max}$ in Fig.\ \ref{fig:critT} appears to be
dominated by the second subleading correction to scaling term.) The exponents resulting
from the fit read $\beta/\nu=0.480(8)$ and $\omega=0.82(2)$. To assess the error arising
from the uncertainty in $T_c$, we repeat the analysis for temperatures $T_c\pm \Delta T_c$
with $\Delta T_c=0.001$.
This leads to shifts of $\beta/\nu$ of about 0.01. Our final estimate therefore reads
$\beta/\nu=0.48(2)$.

\begin{figure}
\includegraphics[width=8.6cm]{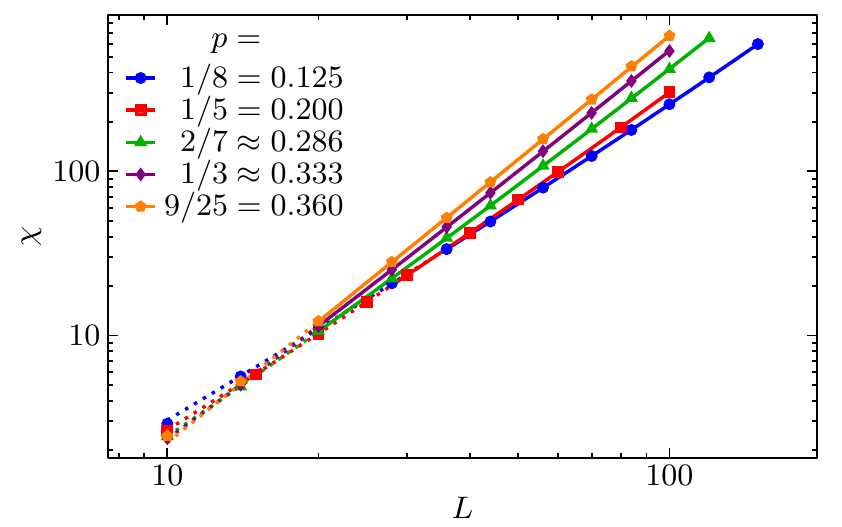}
\caption{(Color online) Double logarithmic plot of $\chi$ vs.\ $L$ for several dilutions $p$ below
        the percolation threshold. Solid lines at fits to $\chi=a L^{\gamma/\nu}(1+b L^{-\omega})$ giving
    $\gamma/\nu=2.524(8)$ and $\omega=0.77(1)$. The lines are dotted in the regions not included
        in the fit. The statistical errors of the data are well below a symbol size. }
\label{fig:scaling_gamma}
\end{figure}

The system-size dependence of the order parameter susceptibility $\chi$ at criticality
is presented in Fig.\  \ref{fig:scaling_gamma} for all dilutions $p<p_c$.
After excluding the smallest system sizes (see figure), the combined fit of all data
to $\chi=a L^{\gamma/\nu}(1+b L^{-\omega})$  is again of good quality ($\tilde \chi^2\approx 1.5$)
and yields the exponents $\gamma/\nu=2.524(8)$ and $\omega=0.77(1)$. After including potential
errors from the uncertainty in $T_c$ and the fit range, the final exponent estimate
is $\gamma/\nu=2.52(4)$.

\begin{figure}
\includegraphics[width=8.6cm]{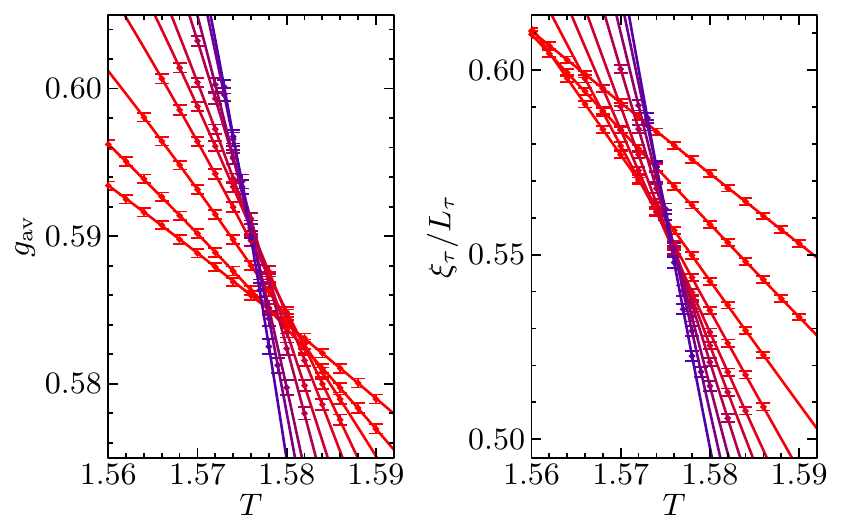}
\caption{(Color online) Average Binder cumulant $g_{\rm av}$ and reduced correlation length $\xi_\tau/L_\tau$
as functions of temperature for dilution $p=1/3$ and systems of optimal shape. System sizes range from
$L=10$ to 100 (as listed in Fig.\ \ref{fig:g_L_Lt}) with increasing slope. }
\label{fig:gT_xitT_03333}
\end{figure}

\begin{figure}
\includegraphics[width=8.6cm]{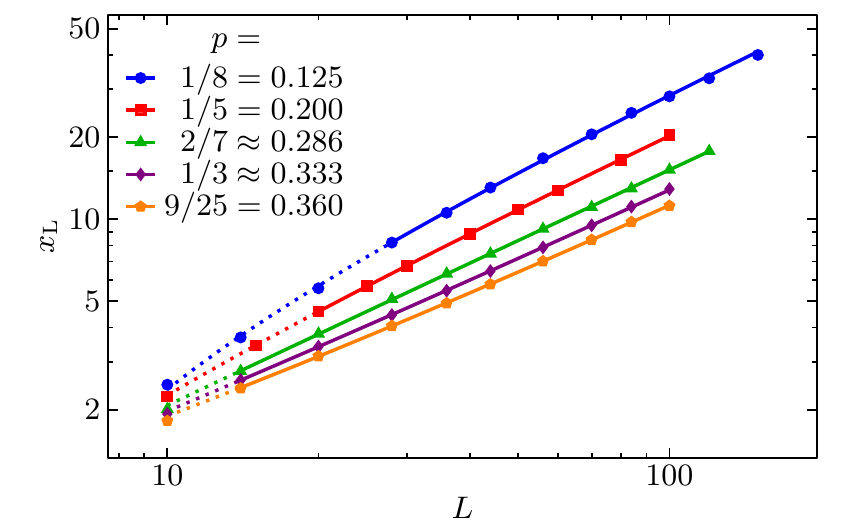}
\caption{(Color online) Slope $x_L = (d/dT)\xi_\tau/L_\tau$ at criticality vs.\
system size $L$ for optimally shaped samples at different dilutions $p$.
Solid lines at fits to $x_L=a L^{1/\nu}(1+b L^{-\omega})$ giving
    $\nu=1.165(6)$ and $\omega=0.74(1)$. The lines are dotted in the regions not included
        in the fit. }
\label{fig:scaling_nu_xit}
\end{figure}

\begin{figure}
\includegraphics[width=8.6cm]{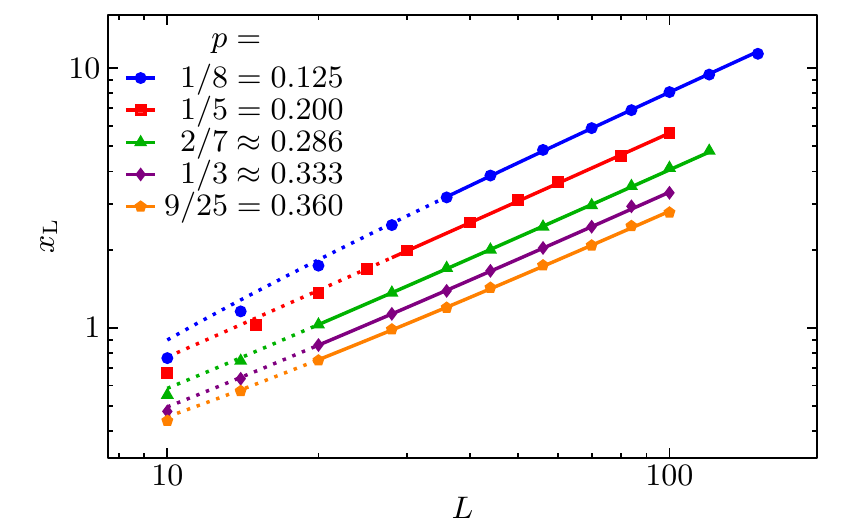}
\caption{(Color online) Slope $x_L = (d/dT)g_{\rm av}$ at criticality vs.\
system size $L$ for optimally shaped samples at different dilutions $p$.
Solid lines at fits to $x_L=aL^{1/\nu}(1+bL^{-\omega})$ giving
    $\nu=1.146(16)$ and $\omega=0.97(23)$. The lines are dotted in the regions not included
        in the fit.}
\label{fig:scaling_nu_g}
\end{figure}

So far, the analysis has focused on the behavior right at $T_c$. To find a complete set of
critical exponents, we now determine the correlation length exponent $\nu$ which requires
off-critical data. Figure \ref{fig:gT_xitT_03333} shows the temperature dependence of the
Binder cumulant $g_{\rm av}$ and the reduced correlation length $\xi_\tau/L_\tau$ for systems
of optimal shape but different sizes at dilution $p=1/3$.
Both quantities have scale dimension zero, therefore, the curves for different system
sizes are expected to cross at the critical temperature $T_c$. The figure demonstrates
that the crossings for both quantities shift with increasing $L$, reflecting significant
corrections to scaling. According to Eqs.\ (\ref{eq:Binder_FSS}) and (\ref{eq:xit}), the slopes
$(d/dT)g_{\rm av}$ and $(d/dT)\xi_\tau/L_\tau$ at the critical temperature $T_c$
vary as $L^{1/\nu}$ with system size. To extract the slopes, we fit straight lines (for $\xi_\tau/L_\tau$)
or quadratic parabolas (for $g_{\rm av}$) to the data close to $T_c$.
The resulting slopes are shown as a function of system size in Figs.
\ref{fig:scaling_nu_xit} and \ref{fig:scaling_nu_g}, respectively.
The exponent $\nu$ is now obtained from fits of the slopes to the form
$aL^{-1/\nu}(1+bL^{-\omega})$. In the case of the reduced correlation length $\xi_\tau/L_\tau$
(Fig.\ \ref{fig:scaling_nu_xit}) a combined fit of all dilutions $p<p_c$ is of good quality
after the smallest system sizes have been excluded ($\tilde \chi^2\approx 1.2)$ and yields
$\nu=1.165(6)$ as well as $\omega=0.74(1)$. The corresponding fit of the slopes of the
Binder cumulant has a somewhat poorer quality ($\tilde \chi^2\approx 5.5)$ and is not very stable with respect to
adding and removing data points at the
ends of the interval. The resulting exponents $\nu=1.146(16)$ and $\omega=0.97(23)$ have
therefore larger errors.
In addition to the slopes of the
Binder cumulant $g_{\rm av}$ and the reduced correlation length $\xi_\tau/L_\tau$ at $T_c$,
we have also studied the slopes of $\xi_s/L$ and $\ln m$ (not shown).
After we account for the differences between all these estimates and
include potential errors from the uncertainty in $T_c$
(by repeating the analysis at temperatures $T_c\pm 0.001$) we arrive at the final estimate
$\nu=1.16(5)$. This value fulfills the inequality \cite{CCFS86}
 $d \nu > 2$.

The critical exponents $\beta/\nu$, $\gamma/\nu$, and $z$ are not independent of each other as
they must fulfill the hyperscaling relation   $2\beta/\nu+\gamma/\nu=d+z$ where
$d=2$ is the space dimensionality. Our values, $\beta/\nu=0.48(2)$, $\gamma/\nu=2.52(4)$, and
$z=1.52(3)$ fulfill this relation within their error bars. We also note that all our estimates
for the leading irrelevant exponent $\omega$ are roughly consistent with each other, giving
us confidence that our results represent true asymptotic rather than effective critical
exponents.

\subsection{Percolation transition}
\label{subsec:MC-Percolation}

We now turn to the percolation transition that occurs when the system is tuned through
the percolation threshold $p_c$ at low (classical) temperatures (see Fig.\ \ref{fig:pd}).
The critical behavior of this transition stems from the critical geometry of the
percolating lattice while the dynamical fluctuations of the rotor variables are
uncritical and ``just go along for the ride''
(the rotor model on each of the percolation clusters is locally ordered).
Vojta and Schmalian \cite{VojtaSchmalian05b} developed a theory of this percolation
quantum phase transition. It predicts critical behavior governed by the lattice percolation
exponents. For two space dimensions it yields the exact exponent values
$\beta=5/36$, $\gamma=59/12$, $\nu=4/3$, and $z=91/48$.

To test these predictions, we perform simulations at dilution $p=p_c=0.407253$
and temperature $T=1.0$. These calculations require a particularly high numerical
effort, because the large value of $z$ leads to a rapid increase with $L$ of the optimal
system size $L_\tau^{\max}$ in imaginary time direction. We have thus restricted the simulations
to sizes up to $L=56$ and $L_\tau=1792$ using between 10\,000 and 50\,000 disorder configurations.

\begin{figure}
\includegraphics[width=8.6cm]{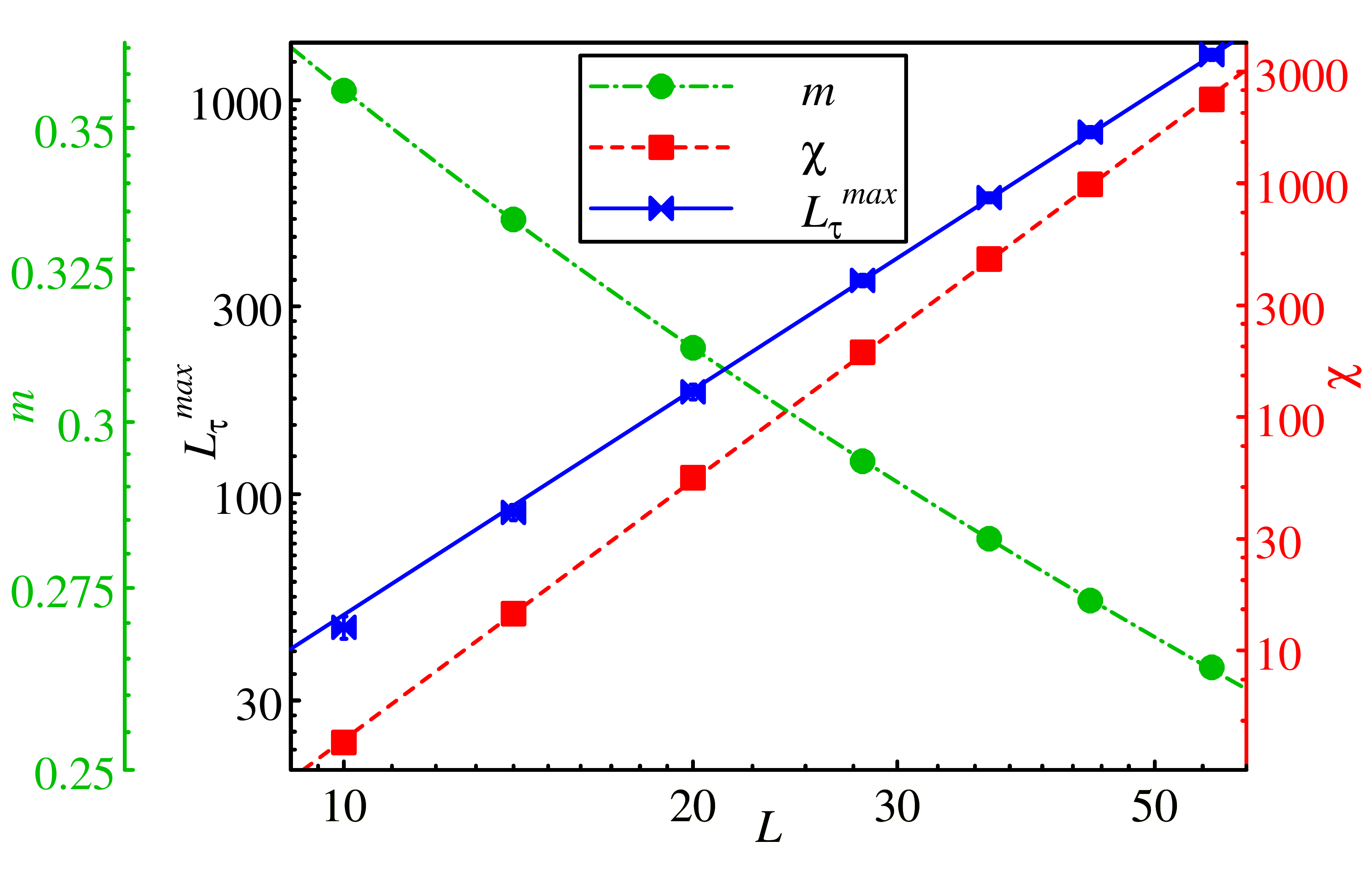}
\caption{(Color online) Double logarithmic plots of $L_\tau^{\rm max}$, $m$ and $\chi$
for dilution $p=p_c=0.407253$ and $T=1.0$. The lines are fits to the predictions of
the Ref.\ \onlinecite{VojtaSchmalian05b}, namely $L_\tau^{\rm max} \sim L^{91/48}$
and $\chi\sim L^{59/16}$. For the order parameter, a subleading correction is included
via $m = aL^{-5/48}(1+bL^{-\omega})$. The statistical errors are of the order of the
symbol size or smaller.}
\label{fig:pc_exponents}
\end{figure}

The data analysis proceeds in analogy to Sec.\ \ref{subsec:MC-Generic}.  We obtain
$L_\tau^{\rm max}$ from the maxima of the Binder cumulant $g_{\rm av}$ as a function of
$L_\tau$ at fixed $L$. In Fig.\ \ref{fig:pc_exponents}, we present a plot
of $L_\tau^{\rm max}$ vs.\ $L$.
The data can be fitted with high quality ($\tilde \chi^2 \approx 0.4$) to the predicted
power law $L_\tau^{\rm max} \sim L^{91/48}$. After having found $L_\tau^{\rm max}$,
we calculate the order parameter and susceptibility right at criticality for
optimally shaped samples of different sizes. The resulting data are also presented
in Fig.\ \ref{fig:pc_exponents}. The susceptibility data can be fitted well to
the predicted power law $\chi \sim L^{59/16}$ giving  $\tilde \chi^2 \approx 0.8$.
The exponent $\beta/\nu=5/48$ is very small, corresponding to
a slow decay of the order parameter $m$ with $L$. Subleading corrections are thus much more
visible as indicated by the curvature of the $m$ vs.\ $L$ curve in Fig.\ \ref{fig:pc_exponents}.
We have therefore fitted the order parameter to $m = aL^{-5/48}(1+bL^{-\omega})$.
This fit is again of high quality, with  $\tilde \chi^2\approx 0.5$.

Our simulation data thus agree nearly perfectly with the critical behavior
predicted in Ref.\ \onlinecite{VojtaSchmalian05b}.

\subsection{Soft-spin model}
\label{subsec:MC-soft}

We also consider a soft-spin version of the classical Hamiltonian to test whether or
not its critical exponents agree with those of the hard-spin model analyzed above,
as is expected from universality. The soft-spin Hamiltonian reads
\begin{eqnarray}
 H_{\rm soft} &=& -\sum_{\langle i,j\rangle, t}
\epsilon_i\epsilon_j\mathbf{S}_{i,t}\cdot\mathbf{S}_{j,t}
- \sum_{i,t}\epsilon_i\mathbf{S}_{i,t}\cdot\mathbf{S}_{i,t+1} \nonumber\\
& & -\frac 1 2 \sum_{i,t}\epsilon_i|\mathbf{S}_{i,t}|^2 + \sum_{i,t}\epsilon_i\left(|\mathbf{S}_{i,t}|^2\right)^2
\label{eq:Hsoft}
\end{eqnarray}
where $\mathbf{S}_{i,t}$ now represents an unrestricted two-component vector.
We perform Monte-Carlo simulations of this soft-spin model using the efficient
Worm algorithm \cite{ProkofevSvistunov01}, studying dilutions $p=0.286$ and 0.337.
The system sizes range from $L=8$ to 24 with $L_\tau$ fixed at $L_\tau=L^z$ using the
dynamical exponent value found
in Sec.\ \ref{subsec:MC-Generic} \footnote{We actually use $z=1.45$ which is
close to the effective dynamical exponent found for the system size range and dilution
considered.}.

\begin{figure}
\includegraphics[width=8.6cm]{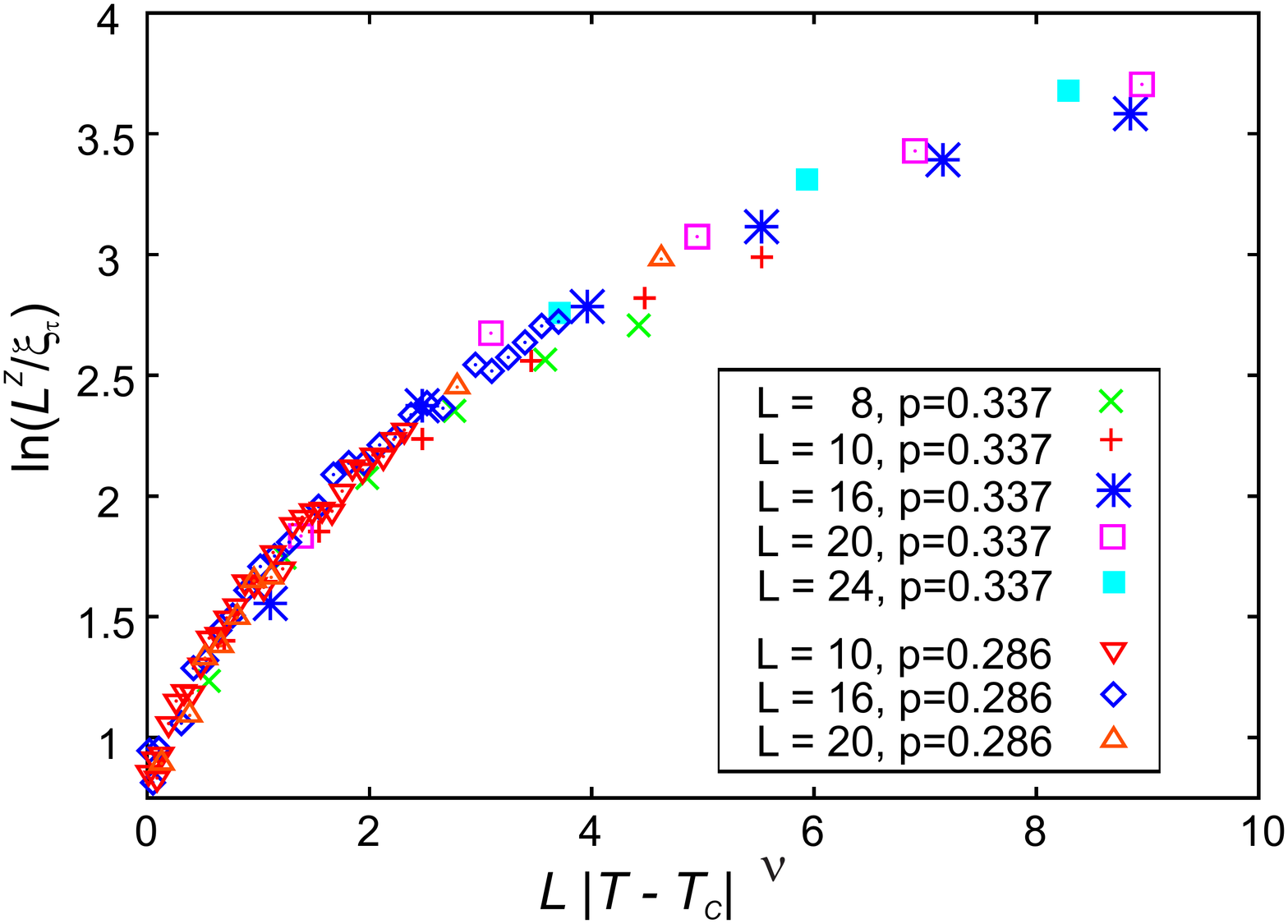}
\caption{(color online) Scaling plot of the correlation length $\xi_\tau$ in imaginary time direction
of the soft-spin model (\ref{eq:Hsoft}). Shown are data for two dilutions $p$, several system sizes $L$, and temperatures $T$ on the disordered side of the transition. The exponents $z$ and $\nu$ are fixed at the values found in Sec.\ \ref{subsec:MC-Generic}. The data are averages over 100 disorder configurations. Their statistical errors are about one symbol size.}
\label{fig:soft}
\end{figure}

We now analyze the correlation length $\xi_\tau$ in imaginary time direction (equivalent
to the inverse energy gap of the corresponding quantum model) on the disordered side
of the phase transition. According to Eq.\ (\ref{eq:xit_FSS}), its scaling form
for samples of shape $L_\tau=L^z$ can be written as $\xi_\tau = L^z X_{\xi_\tau}(rL^{1/\nu},1)$.
Thus, if we plot  $\xi_\tau/L^z$ vs.\ $(T-T_c)L^{1/\nu}$, the data for different sizes and
temperatures should all fall onto a single master curve. Figure \ref{fig:soft}
presents such a plot for two site dilutions $p$, with the critical
exponents $z$ and $\nu$ fixed at the values found in Sec.\ \ref{subsec:MC-Generic}.
Within their statistical errors, the data scale well.
Consequently, even though we have not independently determined the critical exponents
of the soft-spin model  (\ref{eq:Hsoft}), the Monte Carlo data are compatible with
the critical behavior found earlier.

\section{Conclusions}
\label{sec:Conclusions}

In summary, we have carried out large-scale computer simulations to determine
the critical behavior of the superfluid-Mott glass quantum phase transition in two
space dimensions. To this end, we have mapped a quantum rotor model with commensurate
filling and off-diagonal disorder onto a (2+1)-dimensional classical XY model with
columnar defects. We have then analyzed this classical system by means of Monte Carlo methods.

The corresponding clean superfluid-Mott insulator transition is in the three-dimensional
XY universality class; its correlation length exponent $\nu\approx 0.6717$ violates the Harris criterion
$d\nu>2$ with $d=2$. The clean critical behavior is therefore expected to be
unstable against the columnar disorder. Accordingly, we have found that the critical behavior
of the superfluid-Mott glass transition differs from that of the  clean superfluid-Mott
insulator transition.

In contrast to other quantum phase transitions in disordered systems
\cite{Fisher92,*Fisher95,HoyosKotabageVojta07,*VojtaKotabageHoyos09,DRMS08,*DRHV10,Xingetal15,UbaidKassisVojtaSchroeder10,Vojta10},
the superfluid-Mott glass transition features a conventional finite-disorder
critical point whose dynamical scaling is characterized by
a power-law relation $\xi_\tau \sim \xi_s^z$ between the correlation lengths in the
space and time directions
(rather than an infinite-randomness critical point with activated dynamical scaling
for which $\xi_\tau$ would grow exponentially with $\xi_s$).
This result agrees with the general classification
of phase transitions in disordered systems based on the rare region (or defect)
dimensionality \cite{VojtaSchmalian05,Vojta06}. In terms of the mapped, classical
Hamiltonian (\ref{eq:Hcl}), the rare regions in our problem are one-dimensional
rods with XY order-parameter symmetry. As the lower-critical dimension of the
classical XY model is $d_c^-=2$, the rare region dimensionality fulfills
$d_{RR} < d_c^-$, putting the system into the conventional class A of the
classification.

For the generic transition occurring at dilutions $p$ below the lattice percolation
threshold $p_c$, our Monte Carlo data are described well by a universal critical behavior
with dilution-independent critical exponents.
The numerical estimates of the exponent values are summarized in Table
\ref{table:exponents} and compared to earlier results in the literature.
\begin{table}
\renewcommand*{\arraystretch}{1.2}
\begin{tabular*}{\columnwidth}{c @{\extracolsep{\fill}} rrll}
\hline\hline
Value              & This work~    & Ref.\ \onlinecite{ProkofevSvistunov04} & Ref.\ \onlinecite{IyerPekkerRefael12} & Ref.\ \onlinecite{SLRT14} \\
\hline
$\nu$               & 1.16(5)      &                 & 1.09(4)          &  0.96(6)                   \\
$z$                 & 1.52(3)      &  1.5(2)         & 1.31(7)          &  fixed at 1                \\
$\beta/\nu$         & 0.48(2)      &  {\it 0.60(15)} &  {\it 1.1(2)}    &                          \\
$\gamma/\nu$        & 2.52(4)      &  {\it 2.3(1)}   &  1.1(2)          &                              \\
$\eta$              & {\it $-$0.52(4)}   &  $-$0.3(1)      & {\it 0.9(2)}     &                              \\
\hline\hline
\end{tabular*}
\caption{Critical exponents of the superfluid-Mott glass quantum phase transition. Upright numbers are directly given in the respective papers, italic ones were calculated using scaling relations such as $2\beta/\nu+\gamma/\nu=d+z$ and $\eta=2-\gamma/\nu$.
}
\label{table:exponents}
\end{table}
Our results are in reasonable agreement with (but more accurate than) Monte Carlo simulations of a link-current
model \cite{ProkofevSvistunov04} that is expected to be in the same universality class as our Hamiltonian.
The results in Ref.\ \onlinecite{IyerPekkerRefael12} were obtained using a numerical implementation of the
strong-disorder renormalization group. This method is expected to give approximate rather than exact results
at a conventional finite-disorder critical point such as the one under consideration here. In view of this,
the agreement of $\nu$ and $z$ can be considered satisfactory. However, the values of $\beta/\nu$, $\gamma/\nu$,
and $\eta$ (that all involve the scale dimension of the order parameter) are far away from the Monte Carlo results
in this work and in Ref.\ \onlinecite{ProkofevSvistunov04}. Our findings are also incompatible with the clean value
$z=1$ that was assumed in Ref.\  \onlinecite{SLRT14}.

It is interesting to consider the evolution of the dynamical exponent $z$ with the order parameter dimensionality.
The deviation of $z$ from the clean value, which is $z=1$ for any number of components,
can be understood as a measure of the strength of the disorder effects.
In the (2+1)-dimensional Heisenberg model (three order parameter components) with columnar defects, the exponent takes the value \cite{SknepnekVojtaVojta04,VojtaSknepnek06} $z= 1.31$. The (2+1)-dimensional XY model (two components)
studied in the present paper has $z=1.52$, while the corresponding Ising model \cite{MMHF00} (one component)  features activated scaling
that corresponds to $z=\infty$. The value of $z$ thus increases monotonically with decreasing order parameter
dimensionality.

In addition to the generic superfluid-Mott glass transition that occurs for dilutions $p<p_c$, we have also investigated
the percolation quantum phase transition across $p_c$. Here, our Monte Carlo data agree very well with the predictions
of the scaling theory by Vojta and Schmalian \cite{VojtaSchmalian05b}.

Potential routes to study the superfluid-Mott glass transition in experiment include disordered bosonic systems
in ultracold atoms as well as dirty and granualar superconductors (for some superconductor-insulator transitions,
there is experimental and numerical evidence for the bosonic nature of the transition). In these systems,
it may be hard, though, to fulfill the condition of exact particle-hole symmetry in the presence of disorder.
\emph{Statistical} particle hole symmetry may be easier to achieve, but it is not fully resolved whether or not
it would destabilize the Mott glass and turn it into a Bose glass \cite{WeichmanMukhopadhyay08,AKPR10,WangGuoSandvik15}.

Another type of experimental systems that contain Mott-glass physics are diluted anisotropic
spin-1 antiferromagnets \cite{RoscildeHaas07}. In this case, the particle-hole symmetry appears naturally as
it is a consequence of the up-down symmetry of the spin Hamiltonian in the absence of an external magnetic field.
Such a magnetic realization of a Mott glass (albeit in three dimensions)
was recently observed  in bromine-doped dichloro-tetrakis-thiourea-nickel (DTN) \cite{Yuetal12}.

\section*{Acknowledgements}

This work was supported in part by the NSF under Grant Nos.\ DMR-1205803
and DMR-1506152 as well as by funds from the UCSD Academic Senate.
M.P. acknowledges support by an InProTUC scholarship of
the German Academic Exchange Service. We thank Snir Gazit, Gil Refael, and Nandini Trivedi for
helpful discussions.

\bibliographystyle{apsrev4-1}
\bibliography{../00bibtex/rareregions}

\begin{thebibliography}{66}%
\makeatletter
\providecommand \@ifxundefined [1]{%
 \@ifx{#1\undefined}
}%
\providecommand \@ifnum [1]{%
 \ifnum #1\expandafter \@firstoftwo
 \else \expandafter \@secondoftwo
 \fi
}%
\providecommand \@ifx [1]{%
 \ifx #1\expandafter \@firstoftwo
 \else \expandafter \@secondoftwo
 \fi
}%
\providecommand \natexlab [1]{#1}%
\providecommand \enquote  [1]{``#1''}%
\providecommand \bibnamefont  [1]{#1}%
\providecommand \bibfnamefont [1]{#1}%
\providecommand \citenamefont [1]{#1}%
\providecommand \href@noop [0]{\@secondoftwo}%
\providecommand \href [0]{\begingroup \@sanitize@url \@href}%
\providecommand \@href[1]{\@@startlink{#1}\@@href}%
\providecommand \@@href[1]{\endgroup#1\@@endlink}%
\providecommand \@sanitize@url [0]{\catcode `\\12\catcode `\$12\catcode
  `\&12\catcode `\#12\catcode `\^12\catcode `\_12\catcode `\%12\relax}%
\providecommand \@@startlink[1]{}%
\providecommand \@@endlink[0]{}%
\providecommand \url  [0]{\begingroup\@sanitize@url \@url }%
\providecommand \@url [1]{\endgroup\@href {#1}{\urlprefix }}%
\providecommand \urlprefix  [0]{URL }%
\providecommand \Eprint [0]{\href }%
\providecommand \doibase [0]{http://dx.doi.org/}%
\providecommand \selectlanguage [0]{\@gobble}%
\providecommand \bibinfo  [0]{\@secondoftwo}%
\providecommand \bibfield  [0]{\@secondoftwo}%
\providecommand \translation [1]{[#1]}%
\providecommand \BibitemOpen [0]{}%
\providecommand \bibitemStop [0]{}%
\providecommand \bibitemNoStop [0]{.\EOS\space}%
\providecommand \EOS [0]{\spacefactor3000\relax}%
\providecommand \BibitemShut  [1]{\csname bibitem#1\endcsname}%
\let\auto@bib@innerbib\@empty
\bibitem [{\citenamefont {Crooker}\ \emph {et~al.}(1983)\citenamefont
  {Crooker}, \citenamefont {Hebral}, \citenamefont {Smith}, \citenamefont
  {Takano},\ and\ \citenamefont {Reppy}}]{CHSTR83}%
  \BibitemOpen
  \bibfield  {author} {\bibinfo {author} {\bibfnamefont {B.~C.}\ \bibnamefont
  {Crooker}}, \bibinfo {author} {\bibfnamefont {B.}~\bibnamefont {Hebral}},
  \bibinfo {author} {\bibfnamefont {E.~N.}\ \bibnamefont {Smith}}, \bibinfo
  {author} {\bibfnamefont {Y.}~\bibnamefont {Takano}}, \ and\ \bibinfo {author}
  {\bibfnamefont {J.~D.}\ \bibnamefont {Reppy}},\ }\href {\doibase
  10.1103/PhysRevLett.51.666} {\bibfield  {journal} {\bibinfo  {journal} {Phys.
  Rev. Lett.}\ }\textbf {\bibinfo {volume} {51}},\ \bibinfo {pages} {666}
  (\bibinfo {year} {1983})}\BibitemShut {NoStop}%
\bibitem [{\citenamefont {Reppy}(1984)}]{Reppy84}%
  \BibitemOpen
  \bibfield  {author} {\bibinfo {author} {\bibfnamefont {J.~D.}\ \bibnamefont
  {Reppy}},\ }\href@noop {} {\bibfield  {journal} {\bibinfo  {journal} {Physica
  B+C}\ }\textbf {\bibinfo {volume} {126}},\ \bibinfo {pages} {335} (\bibinfo
  {year} {1984})}\BibitemShut {NoStop}%
\bibitem [{\citenamefont {van~der Zant}\ \emph {et~al.}(1992)\citenamefont
  {van~der Zant}, \citenamefont {Fritschy}, \citenamefont {Elion},
  \citenamefont {Geerligs},\ and\ \citenamefont {Mooij}}]{ZFEM92}%
  \BibitemOpen
  \bibfield  {author} {\bibinfo {author} {\bibfnamefont {H.~S.~J.}\
  \bibnamefont {van~der Zant}}, \bibinfo {author} {\bibfnamefont {F.~C.}\
  \bibnamefont {Fritschy}}, \bibinfo {author} {\bibfnamefont {W.~J.}\
  \bibnamefont {Elion}}, \bibinfo {author} {\bibfnamefont {L.~J.}\ \bibnamefont
  {Geerligs}}, \ and\ \bibinfo {author} {\bibfnamefont {J.~E.}\ \bibnamefont
  {Mooij}},\ }\href {\doibase 10.1103/PhysRevLett.69.2971} {\bibfield
  {journal} {\bibinfo  {journal} {Phys. Rev. Lett.}\ }\textbf {\bibinfo
  {volume} {69}},\ \bibinfo {pages} {2971} (\bibinfo {year}
  {1992})}\BibitemShut {NoStop}%
\bibitem [{\citenamefont {van~der Zant}\ \emph {et~al.}(1996)\citenamefont
  {van~der Zant}, \citenamefont {Elion}, \citenamefont {Geerligs},\ and\
  \citenamefont {Mooij}}]{ZEGM96}%
  \BibitemOpen
  \bibfield  {author} {\bibinfo {author} {\bibfnamefont {H.~S.~J.}\
  \bibnamefont {van~der Zant}}, \bibinfo {author} {\bibfnamefont {W.~J.}\
  \bibnamefont {Elion}}, \bibinfo {author} {\bibfnamefont {L.~J.}\ \bibnamefont
  {Geerligs}}, \ and\ \bibinfo {author} {\bibfnamefont {J.~E.}\ \bibnamefont
  {Mooij}},\ }\href {\doibase 10.1103/PhysRevB.54.10081} {\bibfield  {journal}
  {\bibinfo  {journal} {Phys. Rev. B}\ }\textbf {\bibinfo {volume} {54}},\
  \bibinfo {pages} {10081} (\bibinfo {year} {1996})}\BibitemShut {NoStop}%
\bibitem [{\citenamefont {Haviland}\ \emph {et~al.}(1989)\citenamefont
  {Haviland}, \citenamefont {Liu},\ and\ \citenamefont
  {Goldman}}]{HavilandLiuGoldman89}%
  \BibitemOpen
  \bibfield  {author} {\bibinfo {author} {\bibfnamefont {D.~B.}\ \bibnamefont
  {Haviland}}, \bibinfo {author} {\bibfnamefont {Y.}~\bibnamefont {Liu}}, \
  and\ \bibinfo {author} {\bibfnamefont {A.~M.}\ \bibnamefont {Goldman}},\
  }\href {\doibase 10.1103/PhysRevLett.62.2180} {\bibfield  {journal} {\bibinfo
   {journal} {Phys. Rev. Lett.}\ }\textbf {\bibinfo {volume} {62}},\ \bibinfo
  {pages} {2180} (\bibinfo {year} {1989})}\BibitemShut {NoStop}%
\bibitem [{\citenamefont {Hebard}\ and\ \citenamefont
  {Paalanen}(1990)}]{HebardPaalanen90}%
  \BibitemOpen
  \bibfield  {author} {\bibinfo {author} {\bibfnamefont {A.~F.}\ \bibnamefont
  {Hebard}}\ and\ \bibinfo {author} {\bibfnamefont {M.~A.}\ \bibnamefont
  {Paalanen}},\ }\href {\doibase 10.1103/PhysRevLett.65.927} {\bibfield
  {journal} {\bibinfo  {journal} {Phys. Rev. Lett.}\ }\textbf {\bibinfo
  {volume} {65}},\ \bibinfo {pages} {927} (\bibinfo {year} {1990})}\BibitemShut
  {NoStop}%
\bibitem [{\citenamefont {Oosawa}\ and\ \citenamefont
  {Tanaka}(2002)}]{OosawaTanaka02}%
  \BibitemOpen
  \bibfield  {author} {\bibinfo {author} {\bibfnamefont {A.}~\bibnamefont
  {Oosawa}}\ and\ \bibinfo {author} {\bibfnamefont {H.}~\bibnamefont
  {Tanaka}},\ }\href {\doibase 10.1103/PhysRevB.65.184437} {\bibfield
  {journal} {\bibinfo  {journal} {Phys. Rev. B}\ }\textbf {\bibinfo {volume}
  {65}},\ \bibinfo {pages} {184437} (\bibinfo {year} {2002})}\BibitemShut
  {NoStop}%
\bibitem [{\citenamefont {Hong}\ \emph {et~al.}(2010)\citenamefont {Hong},
  \citenamefont {Zheludev}, \citenamefont {Manaka},\ and\ \citenamefont
  {Regnault}}]{HZMR10}%
  \BibitemOpen
  \bibfield  {author} {\bibinfo {author} {\bibfnamefont {T.}~\bibnamefont
  {Hong}}, \bibinfo {author} {\bibfnamefont {A.}~\bibnamefont {Zheludev}},
  \bibinfo {author} {\bibfnamefont {H.}~\bibnamefont {Manaka}}, \ and\ \bibinfo
  {author} {\bibfnamefont {L.-P.}\ \bibnamefont {Regnault}},\ }\href {\doibase
  10.1103/PhysRevB.81.060410} {\bibfield  {journal} {\bibinfo  {journal} {Phys.
  Rev. B}\ }\textbf {\bibinfo {volume} {81}},\ \bibinfo {pages} {060410}
  (\bibinfo {year} {2010})}\BibitemShut {NoStop}%
\bibitem [{\citenamefont {Yu}\ \emph {et~al.}(2012)\citenamefont {Yu},
  \citenamefont {Yin}, \citenamefont {Sullivan}, \citenamefont {Xia},
  \citenamefont {Huan}, \citenamefont {Paduan-Filho}, \citenamefont {Jr},
  \citenamefont {Haas}, \citenamefont {Steppke}, \citenamefont {Miclea},
  \citenamefont {Weickert}, \citenamefont {Movshovich}, \citenamefont {Mun},
  \citenamefont {Scott}, \citenamefont {Zapf},\ and\ \citenamefont
  {Roscilde}}]{Yuetal12}%
  \BibitemOpen
  \bibfield  {author} {\bibinfo {author} {\bibfnamefont {R.}~\bibnamefont
  {Yu}}, \bibinfo {author} {\bibfnamefont {L.}~\bibnamefont {Yin}}, \bibinfo
  {author} {\bibfnamefont {N.~S.}\ \bibnamefont {Sullivan}}, \bibinfo {author}
  {\bibfnamefont {J.~S.}\ \bibnamefont {Xia}}, \bibinfo {author} {\bibfnamefont
  {C.}~\bibnamefont {Huan}}, \bibinfo {author} {\bibfnamefont {A.}~\bibnamefont
  {Paduan-Filho}}, \bibinfo {author} {\bibfnamefont {N.~F.~O.}\ \bibnamefont
  {Jr}}, \bibinfo {author} {\bibfnamefont {S.}~\bibnamefont {Haas}}, \bibinfo
  {author} {\bibfnamefont {A.}~\bibnamefont {Steppke}}, \bibinfo {author}
  {\bibfnamefont {C.~F.}\ \bibnamefont {Miclea}}, \bibinfo {author}
  {\bibfnamefont {F.}~\bibnamefont {Weickert}}, \bibinfo {author}
  {\bibfnamefont {R.}~\bibnamefont {Movshovich}}, \bibinfo {author}
  {\bibfnamefont {E.-D.}\ \bibnamefont {Mun}}, \bibinfo {author} {\bibfnamefont
  {B.~L.}\ \bibnamefont {Scott}}, \bibinfo {author} {\bibfnamefont {V.~S.}\
  \bibnamefont {Zapf}}, \ and\ \bibinfo {author} {\bibfnamefont
  {T.}~\bibnamefont {Roscilde}},\ }\href@noop {} {\bibfield  {journal}
  {\bibinfo  {journal} {Nature}\ }\textbf {\bibinfo {volume} {489}},\ \bibinfo
  {pages} {379} (\bibinfo {year} {2012})}\BibitemShut {NoStop}%
\bibitem [{\citenamefont {White}\ \emph {et~al.}(2009)\citenamefont {White},
  \citenamefont {Pasienski}, \citenamefont {McKay}, \citenamefont {Zhou},
  \citenamefont {Ceperley},\ and\ \citenamefont {DeMarco}}]{WPMZCD09}%
  \BibitemOpen
  \bibfield  {author} {\bibinfo {author} {\bibfnamefont {M.}~\bibnamefont
  {White}}, \bibinfo {author} {\bibfnamefont {M.}~\bibnamefont {Pasienski}},
  \bibinfo {author} {\bibfnamefont {D.}~\bibnamefont {McKay}}, \bibinfo
  {author} {\bibfnamefont {S.~Q.}\ \bibnamefont {Zhou}}, \bibinfo {author}
  {\bibfnamefont {D.}~\bibnamefont {Ceperley}}, \ and\ \bibinfo {author}
  {\bibfnamefont {B.}~\bibnamefont {DeMarco}},\ }\href {\doibase
  10.1103/PhysRevLett.102.055301} {\bibfield  {journal} {\bibinfo  {journal}
  {Phys. Rev. Lett.}\ }\textbf {\bibinfo {volume} {102}},\ \bibinfo {pages}
  {055301} (\bibinfo {year} {2009})}\BibitemShut {NoStop}%
\bibitem [{\citenamefont {Krinner}\ \emph {et~al.}(2013)\citenamefont
  {Krinner}, \citenamefont {Stadler}, \citenamefont {Meineke}, \citenamefont
  {Brantut},\ and\ \citenamefont {Esslinger}}]{KSMBE13}%
  \BibitemOpen
  \bibfield  {author} {\bibinfo {author} {\bibfnamefont {S.}~\bibnamefont
  {Krinner}}, \bibinfo {author} {\bibfnamefont {D.}~\bibnamefont {Stadler}},
  \bibinfo {author} {\bibfnamefont {J.}~\bibnamefont {Meineke}}, \bibinfo
  {author} {\bibfnamefont {J.-P.}\ \bibnamefont {Brantut}}, \ and\ \bibinfo
  {author} {\bibfnamefont {T.}~\bibnamefont {Esslinger}},\ }\href {\doibase
  10.1103/PhysRevLett.110.100601} {\bibfield  {journal} {\bibinfo  {journal}
  {Phys. Rev. Lett.}\ }\textbf {\bibinfo {volume} {110}},\ \bibinfo {pages}
  {100601} (\bibinfo {year} {2013})}\BibitemShut {NoStop}%
\bibitem [{\citenamefont {D'Errico}\ \emph {et~al.}(2014)\citenamefont
  {D'Errico}, \citenamefont {Lucioni}, \citenamefont {Tanzi}, \citenamefont
  {Gori}, \citenamefont {Roux}, \citenamefont {McCulloch}, \citenamefont
  {Giamarchi}, \citenamefont {Inguscio},\ and\ \citenamefont
  {Modugno}}]{DTGRMGIM14}%
  \BibitemOpen
  \bibfield  {author} {\bibinfo {author} {\bibfnamefont {C.}~\bibnamefont
  {D'Errico}}, \bibinfo {author} {\bibfnamefont {E.}~\bibnamefont {Lucioni}},
  \bibinfo {author} {\bibfnamefont {L.}~\bibnamefont {Tanzi}}, \bibinfo
  {author} {\bibfnamefont {L.}~\bibnamefont {Gori}}, \bibinfo {author}
  {\bibfnamefont {G.}~\bibnamefont {Roux}}, \bibinfo {author} {\bibfnamefont
  {I.~P.}\ \bibnamefont {McCulloch}}, \bibinfo {author} {\bibfnamefont
  {T.}~\bibnamefont {Giamarchi}}, \bibinfo {author} {\bibfnamefont
  {M.}~\bibnamefont {Inguscio}}, \ and\ \bibinfo {author} {\bibfnamefont
  {G.}~\bibnamefont {Modugno}},\ }\href {\doibase
  10.1103/PhysRevLett.113.095301} {\bibfield  {journal} {\bibinfo  {journal}
  {Phys. Rev. Lett.}\ }\textbf {\bibinfo {volume} {113}},\ \bibinfo {pages}
  {095301} (\bibinfo {year} {2014})}\BibitemShut {NoStop}%
\bibitem [{\citenamefont {Fisher}\ and\ \citenamefont
  {Fisher}(1988)}]{FisherFisher88}%
  \BibitemOpen
  \bibfield  {author} {\bibinfo {author} {\bibfnamefont {D.~S.}\ \bibnamefont
  {Fisher}}\ and\ \bibinfo {author} {\bibfnamefont {M.~P.~A.}\ \bibnamefont
  {Fisher}},\ }\href {\doibase 10.1103/PhysRevLett.61.1847} {\bibfield
  {journal} {\bibinfo  {journal} {Phys. Rev. Lett.}\ }\textbf {\bibinfo
  {volume} {61}},\ \bibinfo {pages} {1847} (\bibinfo {year}
  {1988})}\BibitemShut {NoStop}%
\bibitem [{\citenamefont {Fisher}\ \emph {et~al.}(1989)\citenamefont {Fisher},
  \citenamefont {Weichman}, \citenamefont {Grinstein},\ and\ \citenamefont
  {Fisher}}]{FWGF89}%
  \BibitemOpen
  \bibfield  {author} {\bibinfo {author} {\bibfnamefont {M.~P.~A.}\
  \bibnamefont {Fisher}}, \bibinfo {author} {\bibfnamefont {P.~B.}\
  \bibnamefont {Weichman}}, \bibinfo {author} {\bibfnamefont {G.}~\bibnamefont
  {Grinstein}}, \ and\ \bibinfo {author} {\bibfnamefont {D.~S.}\ \bibnamefont
  {Fisher}},\ }\href {\doibase 10.1103/PhysRevB.40.546} {\bibfield  {journal}
  {\bibinfo  {journal} {Phys. Rev. B}\ }\textbf {\bibinfo {volume} {40}},\
  \bibinfo {pages} {546} (\bibinfo {year} {1989})}\BibitemShut {NoStop}%
\bibitem [{\citenamefont {Pollet}\ \emph {et~al.}(2009)\citenamefont {Pollet},
  \citenamefont {Prokof'ev}, \citenamefont {Svistunov},\ and\ \citenamefont
  {Troyer}}]{PPST09}%
  \BibitemOpen
  \bibfield  {author} {\bibinfo {author} {\bibfnamefont {L.}~\bibnamefont
  {Pollet}}, \bibinfo {author} {\bibfnamefont {N.~V.}\ \bibnamefont
  {Prokof'ev}}, \bibinfo {author} {\bibfnamefont {B.~V.}\ \bibnamefont
  {Svistunov}}, \ and\ \bibinfo {author} {\bibfnamefont {M.}~\bibnamefont
  {Troyer}},\ }\href {\doibase 10.1103/PhysRevLett.103.140402} {\bibfield
  {journal} {\bibinfo  {journal} {Phys. Rev. Lett.}\ }\textbf {\bibinfo
  {volume} {103}},\ \bibinfo {pages} {140402} (\bibinfo {year}
  {2009})}\BibitemShut {NoStop}%
\bibitem [{\citenamefont {Griffiths}(1969)}]{Griffiths69}%
  \BibitemOpen
  \bibfield  {author} {\bibinfo {author} {\bibfnamefont {R.~B.}\ \bibnamefont
  {Griffiths}},\ }\href {\doibase 10.1103/PhysRevLett.23.17} {\bibfield
  {journal} {\bibinfo  {journal} {Phys. Rev. Lett.}\ }\textbf {\bibinfo
  {volume} {23}},\ \bibinfo {pages} {17} (\bibinfo {year} {1969})}\BibitemShut
  {NoStop}%
\bibitem [{\citenamefont {Thill}\ and\ \citenamefont
  {Huse}(1995)}]{ThillHuse95}%
  \BibitemOpen
  \bibfield  {author} {\bibinfo {author} {\bibfnamefont {M.}~\bibnamefont
  {Thill}}\ and\ \bibinfo {author} {\bibfnamefont {D.~A.}\ \bibnamefont
  {Huse}},\ }\href {\doibase 10.1016/0378-4371(94)00247-Q} {\bibfield
  {journal} {\bibinfo  {journal} {Physica A}\ }\textbf {\bibinfo {volume}
  {214}},\ \bibinfo {pages} {321} (\bibinfo {year} {1995})}\BibitemShut
  {NoStop}%
\bibitem [{\citenamefont {Vojta}(2006)}]{Vojta06}%
  \BibitemOpen
  \bibfield  {author} {\bibinfo {author} {\bibfnamefont {T.}~\bibnamefont
  {Vojta}},\ }\href {\doibase 10.1088/0305-4470/39/22/R01} {\bibfield
  {journal} {\bibinfo  {journal} {J. Phys. A}\ }\textbf {\bibinfo {volume}
  {39}},\ \bibinfo {pages} {R143} (\bibinfo {year} {2006})}\BibitemShut
  {NoStop}%
\bibitem [{\citenamefont {Weichman}\ and\ \citenamefont
  {Mukhopadhyay}(2007)}]{WeichmanMukhopadhyay07}%
  \BibitemOpen
  \bibfield  {author} {\bibinfo {author} {\bibfnamefont {P.~B.}\ \bibnamefont
  {Weichman}}\ and\ \bibinfo {author} {\bibfnamefont {R.}~\bibnamefont
  {Mukhopadhyay}},\ }\href {\doibase 10.1103/PhysRevLett.98.245701} {\bibfield
  {journal} {\bibinfo  {journal} {Phys. Rev. Lett.}\ }\textbf {\bibinfo
  {volume} {98}},\ \bibinfo {pages} {245701} (\bibinfo {year}
  {2007})}\BibitemShut {NoStop}%
\bibitem [{\citenamefont {Priyadarshee}\ \emph {et~al.}(2006)\citenamefont
  {Priyadarshee}, \citenamefont {Chandrasekharan}, \citenamefont {Lee},\ and\
  \citenamefont {Baranger}}]{PCLB06}%
  \BibitemOpen
  \bibfield  {author} {\bibinfo {author} {\bibfnamefont {A.}~\bibnamefont
  {Priyadarshee}}, \bibinfo {author} {\bibfnamefont {S.}~\bibnamefont
  {Chandrasekharan}}, \bibinfo {author} {\bibfnamefont {J.-W.}\ \bibnamefont
  {Lee}}, \ and\ \bibinfo {author} {\bibfnamefont {H.~U.}\ \bibnamefont
  {Baranger}},\ }\href {\doibase 10.1103/PhysRevLett.97.115703} {\bibfield
  {journal} {\bibinfo  {journal} {Phys. Rev. Lett.}\ }\textbf {\bibinfo
  {volume} {97}},\ \bibinfo {pages} {115703} (\bibinfo {year}
  {2006})}\BibitemShut {NoStop}%
\bibitem [{\citenamefont {Meier}\ and\ \citenamefont
  {Wallin}(2012)}]{MeierWallin12}%
  \BibitemOpen
  \bibfield  {author} {\bibinfo {author} {\bibfnamefont {H.}~\bibnamefont
  {Meier}}\ and\ \bibinfo {author} {\bibfnamefont {M.}~\bibnamefont {Wallin}},\
  }\href {\doibase 10.1103/PhysRevLett.108.055701} {\bibfield  {journal}
  {\bibinfo  {journal} {Phys. Rev. Lett.}\ }\textbf {\bibinfo {volume} {108}},\
  \bibinfo {pages} {055701} (\bibinfo {year} {2012})}\BibitemShut {NoStop}%
\bibitem [{\citenamefont {Ng}\ and\ \citenamefont
  {S\o{}rensen}(2015)}]{NgSorensen15}%
  \BibitemOpen
  \bibfield  {author} {\bibinfo {author} {\bibfnamefont {R.}~\bibnamefont
  {Ng}}\ and\ \bibinfo {author} {\bibfnamefont {E.~S.}\ \bibnamefont
  {S\o{}rensen}},\ }\href {\doibase 10.1103/PhysRevLett.114.255701} {\bibfield
  {journal} {\bibinfo  {journal} {Phys. Rev. Lett.}\ }\textbf {\bibinfo
  {volume} {114}},\ \bibinfo {pages} {255701} (\bibinfo {year}
  {2015})}\BibitemShut {NoStop}%
\bibitem [{\citenamefont {\'Alvarez Z\'u\~niga}\ \emph
  {et~al.}(2015)\citenamefont {\'Alvarez Z\'u\~niga}, \citenamefont {Luitz},
  \citenamefont {Lemari\'e},\ and\ \citenamefont {Laflorencie}}]{ALLL15}%
  \BibitemOpen
  \bibfield  {author} {\bibinfo {author} {\bibfnamefont {J.~P.}\ \bibnamefont
  {\'Alvarez Z\'u\~niga}}, \bibinfo {author} {\bibfnamefont {D.~J.}\
  \bibnamefont {Luitz}}, \bibinfo {author} {\bibfnamefont {G.}~\bibnamefont
  {Lemari\'e}}, \ and\ \bibinfo {author} {\bibfnamefont {N.}~\bibnamefont
  {Laflorencie}},\ }\href {\doibase 10.1103/PhysRevLett.114.155301} {\bibfield
  {journal} {\bibinfo  {journal} {Phys. Rev. Lett.}\ }\textbf {\bibinfo
  {volume} {114}},\ \bibinfo {pages} {155301} (\bibinfo {year}
  {2015})}\BibitemShut {NoStop}%
\bibitem [{\citenamefont {Giamarchi}\ \emph {et~al.}(2001)\citenamefont
  {Giamarchi}, \citenamefont {Le~Doussal},\ and\ \citenamefont
  {Orignac}}]{GiamarchiLeDoussalOrignac01}%
  \BibitemOpen
  \bibfield  {author} {\bibinfo {author} {\bibfnamefont {T.}~\bibnamefont
  {Giamarchi}}, \bibinfo {author} {\bibfnamefont {P.}~\bibnamefont
  {Le~Doussal}}, \ and\ \bibinfo {author} {\bibfnamefont {E.}~\bibnamefont
  {Orignac}},\ }\href {\doibase 10.1103/PhysRevB.64.245119} {\bibfield
  {journal} {\bibinfo  {journal} {Phys. Rev. B}\ }\textbf {\bibinfo {volume}
  {64}},\ \bibinfo {pages} {245119} (\bibinfo {year} {2001})}\BibitemShut
  {NoStop}%
\bibitem [{\citenamefont {Weichman}\ and\ \citenamefont
  {Mukhopadhyay}(2008)}]{WeichmanMukhopadhyay08}%
  \BibitemOpen
  \bibfield  {author} {\bibinfo {author} {\bibfnamefont {P.~B.}\ \bibnamefont
  {Weichman}}\ and\ \bibinfo {author} {\bibfnamefont {R.}~\bibnamefont
  {Mukhopadhyay}},\ }\href {\doibase 10.1103/PhysRevB.77.214516} {\bibfield
  {journal} {\bibinfo  {journal} {Phys. Rev. B}\ }\textbf {\bibinfo {volume}
  {77}},\ \bibinfo {pages} {214516} (\bibinfo {year} {2008})}\BibitemShut
  {NoStop}%
\bibitem [{\citenamefont {Prokof'ev}\ and\ \citenamefont
  {Svistunov}(2004)}]{ProkofevSvistunov04}%
  \BibitemOpen
  \bibfield  {author} {\bibinfo {author} {\bibfnamefont {N.}~\bibnamefont
  {Prokof'ev}}\ and\ \bibinfo {author} {\bibfnamefont {B.}~\bibnamefont
  {Svistunov}},\ }\href {\doibase 10.1103/PhysRevLett.92.015703} {\bibfield
  {journal} {\bibinfo  {journal} {Phys. Rev. Lett.}\ }\textbf {\bibinfo
  {volume} {92}},\ \bibinfo {pages} {015703} (\bibinfo {year}
  {2004})}\BibitemShut {NoStop}%
\bibitem [{Note1()}]{Note1}%
  \BibitemOpen
  \bibinfo {note} {Here, the numbers in parentheses are the errors of the last
  digits.}\BibitemShut {Stop}%
\bibitem [{\citenamefont {Iyer}\ \emph {et~al.}(2012)\citenamefont {Iyer},
  \citenamefont {Pekker},\ and\ \citenamefont {Refael}}]{IyerPekkerRefael12}%
  \BibitemOpen
  \bibfield  {author} {\bibinfo {author} {\bibfnamefont {S.}~\bibnamefont
  {Iyer}}, \bibinfo {author} {\bibfnamefont {D.}~\bibnamefont {Pekker}}, \ and\
  \bibinfo {author} {\bibfnamefont {G.}~\bibnamefont {Refael}},\ }\href
  {\doibase 10.1103/PhysRevB.85.094202} {\bibfield  {journal} {\bibinfo
  {journal} {Phys. Rev. B}\ }\textbf {\bibinfo {volume} {85}},\ \bibinfo
  {pages} {094202} (\bibinfo {year} {2012})}\BibitemShut {NoStop}%
\bibitem [{\citenamefont {Swanson}\ \emph {et~al.}(2014)\citenamefont
  {Swanson}, \citenamefont {Loh}, \citenamefont {Randeria},\ and\ \citenamefont
  {Trivedi}}]{SLRT14}%
  \BibitemOpen
  \bibfield  {author} {\bibinfo {author} {\bibfnamefont {M.}~\bibnamefont
  {Swanson}}, \bibinfo {author} {\bibfnamefont {Y.~L.}\ \bibnamefont {Loh}},
  \bibinfo {author} {\bibfnamefont {M.}~\bibnamefont {Randeria}}, \ and\
  \bibinfo {author} {\bibfnamefont {N.}~\bibnamefont {Trivedi}},\ }\href
  {\doibase 10.1103/PhysRevX.4.021007} {\bibfield  {journal} {\bibinfo
  {journal} {Phys. Rev. X}\ }\textbf {\bibinfo {volume} {4}},\ \bibinfo {pages}
  {021007} (\bibinfo {year} {2014})}\BibitemShut {NoStop}%
\bibitem [{\citenamefont {Guo}\ \emph {et~al.}(1994)\citenamefont {Guo},
  \citenamefont {Bhatt},\ and\ \citenamefont {Huse}}]{GuoBhattHuse94}%
  \BibitemOpen
  \bibfield  {author} {\bibinfo {author} {\bibfnamefont {M.}~\bibnamefont
  {Guo}}, \bibinfo {author} {\bibfnamefont {R.~N.}\ \bibnamefont {Bhatt}}, \
  and\ \bibinfo {author} {\bibfnamefont {D.~A.}\ \bibnamefont {Huse}},\
  }\href@noop {} {\bibfield  {journal} {\bibinfo  {journal} {Phys. Rev. Lett.}\
  }\textbf {\bibinfo {volume} {72}},\ \bibinfo {pages} {4137} (\bibinfo {year}
  {1994})}\BibitemShut {NoStop}%
\bibitem [{\citenamefont {Rieger}\ and\ \citenamefont
  {Young}(1994)}]{RiegerYoung94}%
  \BibitemOpen
  \bibfield  {author} {\bibinfo {author} {\bibfnamefont {H.}~\bibnamefont
  {Rieger}}\ and\ \bibinfo {author} {\bibfnamefont {A.~P.}\ \bibnamefont
  {Young}},\ }\href@noop {} {\bibfield  {journal} {\bibinfo  {journal} {Phys.
  Rev. Lett.}\ }\textbf {\bibinfo {volume} {72}},\ \bibinfo {pages} {4141}
  (\bibinfo {year} {1994})}\BibitemShut {NoStop}%
\bibitem [{\citenamefont {Sknepnek}\ \emph {et~al.}(2004)\citenamefont
  {Sknepnek}, \citenamefont {Vojta},\ and\ \citenamefont
  {Vojta}}]{SknepnekVojtaVojta04}%
  \BibitemOpen
  \bibfield  {author} {\bibinfo {author} {\bibfnamefont {R.}~\bibnamefont
  {Sknepnek}}, \bibinfo {author} {\bibfnamefont {T.}~\bibnamefont {Vojta}}, \
  and\ \bibinfo {author} {\bibfnamefont {M.}~\bibnamefont {Vojta}},\
  }\href@noop {} {\bibfield  {journal} {\bibinfo  {journal} {Phys. Rev. Lett.}\
  }\textbf {\bibinfo {volume} {93}},\ \bibinfo {pages} {097201} (\bibinfo
  {year} {2004})}\BibitemShut {NoStop}%
\bibitem [{\citenamefont {Vojta}\ and\ \citenamefont
  {Sknepnek}(2006)}]{VojtaSknepnek06}%
  \BibitemOpen
  \bibfield  {author} {\bibinfo {author} {\bibfnamefont {T.}~\bibnamefont
  {Vojta}}\ and\ \bibinfo {author} {\bibfnamefont {R.}~\bibnamefont
  {Sknepnek}},\ }\href@noop {} {\bibfield  {journal} {\bibinfo  {journal}
  {Phys. Rev. B.}\ }\textbf {\bibinfo {volume} {74}},\ \bibinfo {pages}
  {094415} (\bibinfo {year} {2006})}\BibitemShut {NoStop}%
\bibitem [{\citenamefont {Vojta}\ and\ \citenamefont
  {Schmalian}(2005{\natexlab{a}})}]{VojtaSchmalian05b}%
  \BibitemOpen
  \bibfield  {author} {\bibinfo {author} {\bibfnamefont {T.}~\bibnamefont
  {Vojta}}\ and\ \bibinfo {author} {\bibfnamefont {J.}~\bibnamefont
  {Schmalian}},\ }\href@noop {} {\bibfield  {journal} {\bibinfo  {journal}
  {Phys. Rev. Lett.}\ }\textbf {\bibinfo {volume} {95}},\ \bibinfo {pages}
  {237206} (\bibinfo {year} {2005}{\natexlab{a}})}\BibitemShut {NoStop}%
\bibitem [{\citenamefont {Wallin}\ \emph {et~al.}(1994)\citenamefont {Wallin},
  \citenamefont {Sorensen}, \citenamefont {Girvin},\ and\ \citenamefont
  {Young}}]{WSGY94}%
  \BibitemOpen
  \bibfield  {author} {\bibinfo {author} {\bibfnamefont {M.}~\bibnamefont
  {Wallin}}, \bibinfo {author} {\bibfnamefont {E.~S.}\ \bibnamefont
  {Sorensen}}, \bibinfo {author} {\bibfnamefont {S.~M.}\ \bibnamefont
  {Girvin}}, \ and\ \bibinfo {author} {\bibfnamefont {A.~P.}\ \bibnamefont
  {Young}},\ }\href@noop {} {\bibfield  {journal} {\bibinfo  {journal} {Phys.
  Rev. B}\ }\textbf {\bibinfo {volume} {49}},\ \bibinfo {pages} {12115}
  (\bibinfo {year} {1994})}\BibitemShut {NoStop}%
\bibitem [{\citenamefont {Campostrini}\ \emph {et~al.}(2006)\citenamefont
  {Campostrini}, \citenamefont {Hasenbusch}, \citenamefont {Pelissetto},\ and\
  \citenamefont {Vicari}}]{CHPV06}%
  \BibitemOpen
  \bibfield  {author} {\bibinfo {author} {\bibfnamefont {M.}~\bibnamefont
  {Campostrini}}, \bibinfo {author} {\bibfnamefont {M.}~\bibnamefont
  {Hasenbusch}}, \bibinfo {author} {\bibfnamefont {A.}~\bibnamefont
  {Pelissetto}}, \ and\ \bibinfo {author} {\bibfnamefont {E.}~\bibnamefont
  {Vicari}},\ }\href {\doibase 10.1103/PhysRevB.74.144506} {\bibfield
  {journal} {\bibinfo  {journal} {Phys. Rev. B}\ }\textbf {\bibinfo {volume}
  {74}},\ \bibinfo {pages} {144506} (\bibinfo {year} {2006})}\BibitemShut
  {NoStop}%
\bibitem [{\citenamefont {Harris}(1974)}]{Harris74}%
  \BibitemOpen
  \bibfield  {author} {\bibinfo {author} {\bibfnamefont {A.~B.}\ \bibnamefont
  {Harris}},\ }\href {\doibase 10.1088/0022-3719/7/9/009} {\bibfield  {journal}
  {\bibinfo  {journal} {J. Phys. C}\ }\textbf {\bibinfo {volume} {7}},\
  \bibinfo {pages} {1671} (\bibinfo {year} {1974})}\BibitemShut {NoStop}%
\bibitem [{\citenamefont {Barber}(1983)}]{Barber_review83}%
  \BibitemOpen
  \bibfield  {author} {\bibinfo {author} {\bibfnamefont {M.~N.}\ \bibnamefont
  {Barber}},\ }in\ \href@noop {} {\emph {\bibinfo {booktitle} {Phase
  Transitions and Critical Phenomena}}},\ Vol.~\bibinfo {volume} {8},\ \bibinfo
  {editor} {edited by\ \bibinfo {editor} {\bibfnamefont {C.}~\bibnamefont
  {Domb}}\ and\ \bibinfo {editor} {\bibfnamefont {J.~L.}\ \bibnamefont
  {Lebowitz}}}\ (\bibinfo  {publisher} {Academic},\ \bibinfo {address} {New
  York},\ \bibinfo {year} {1983})\ pp.\ \bibinfo {pages} {145--266}\BibitemShut
  {NoStop}%
\bibitem [{\citenamefont {Cardy}(1988)}]{Cardy_book88}%
  \BibitemOpen
  \bibinfo {editor} {\bibfnamefont {J.}~\bibnamefont {Cardy}},\ ed.,\
  \href@noop {} {\emph {\bibinfo {title} {Finite-size scaling}}}\ (\bibinfo
  {publisher} {North Holland},\ \bibinfo {address} {Amsterdam},\ \bibinfo
  {year} {1988})\BibitemShut {NoStop}%
\bibitem [{\citenamefont {Fisher}(1992)}]{Fisher92}%
  \BibitemOpen
  \bibfield  {author} {\bibinfo {author} {\bibfnamefont {D.~S.}\ \bibnamefont
  {Fisher}},\ }\href@noop {} {\bibfield  {journal} {\bibinfo  {journal} {Phys.
  Rev. Lett.}\ }\textbf {\bibinfo {volume} {69}},\ \bibinfo {pages} {534}
  (\bibinfo {year} {1992})}\BibitemShut {NoStop}%
\bibitem [{\citenamefont {Fisher}(1995)}]{Fisher95}%
  \BibitemOpen
  \bibfield  {author} {\bibinfo {author} {\bibfnamefont {D.~S.}\ \bibnamefont
  {Fisher}},\ }\href {\doibase 10.1103/PhysRevB.51.6411} {\bibfield  {journal}
  {\bibinfo  {journal} {Phys. Rev. B}\ }\textbf {\bibinfo {volume} {51}},\
  \bibinfo {pages} {6411} (\bibinfo {year} {1995})}\BibitemShut {NoStop}%
\bibitem [{\citenamefont {Hoyos}\ \emph {et~al.}(2007)\citenamefont {Hoyos},
  \citenamefont {Kotabage},\ and\ \citenamefont
  {Vojta}}]{HoyosKotabageVojta07}%
  \BibitemOpen
  \bibfield  {author} {\bibinfo {author} {\bibfnamefont {J.~A.}\ \bibnamefont
  {Hoyos}}, \bibinfo {author} {\bibfnamefont {C.}~\bibnamefont {Kotabage}}, \
  and\ \bibinfo {author} {\bibfnamefont {T.}~\bibnamefont {Vojta}},\ }\href
  {\doibase 10.1103/PhysRevLett.99.230601} {\bibfield  {journal} {\bibinfo
  {journal} {Phys. Rev. Lett.}\ }\textbf {\bibinfo {volume} {99}},\ \bibinfo
  {pages} {230601} (\bibinfo {year} {2007})}\BibitemShut {NoStop}%
\bibitem [{\citenamefont {Vojta}\ \emph {et~al.}(2009)\citenamefont {Vojta},
  \citenamefont {Kotabage},\ and\ \citenamefont
  {Hoyos}}]{VojtaKotabageHoyos09}%
  \BibitemOpen
  \bibfield  {author} {\bibinfo {author} {\bibfnamefont {T.}~\bibnamefont
  {Vojta}}, \bibinfo {author} {\bibfnamefont {C.}~\bibnamefont {Kotabage}}, \
  and\ \bibinfo {author} {\bibfnamefont {J.~A.}\ \bibnamefont {Hoyos}},\ }\href
  {\doibase 10.1103/PhysRevB.79.024401} {\bibfield  {journal} {\bibinfo
  {journal} {Phys. Rev. B}\ }\textbf {\bibinfo {volume} {79}},\ \bibinfo
  {pages} {024401} (\bibinfo {year} {2009})}\BibitemShut {NoStop}%
\bibitem [{\citenamefont {Del~Maestro}\ \emph {et~al.}(2008)\citenamefont
  {Del~Maestro}, \citenamefont {Rosenow}, \citenamefont {M{\"u}ller},\ and\
  \citenamefont {Sachdev}}]{DRMS08}%
  \BibitemOpen
  \bibfield  {author} {\bibinfo {author} {\bibfnamefont {A.}~\bibnamefont
  {Del~Maestro}}, \bibinfo {author} {\bibfnamefont {B.}~\bibnamefont
  {Rosenow}}, \bibinfo {author} {\bibfnamefont {M.}~\bibnamefont {M{\"u}ller}},
  \ and\ \bibinfo {author} {\bibfnamefont {S.}~\bibnamefont {Sachdev}},\
  }\href@noop {} {\bibfield  {journal} {\bibinfo  {journal} {Phys. Rev. Lett.}\
  }\textbf {\bibinfo {volume} {101}},\ \bibinfo {pages} {035701} (\bibinfo
  {year} {2008})}\BibitemShut {NoStop}%
\bibitem [{\citenamefont {Del~Maestro}\ \emph {et~al.}(2010)\citenamefont
  {Del~Maestro}, \citenamefont {Rosenow}, \citenamefont {Hoyos},\ and\
  \citenamefont {Vojta}}]{DRHV10}%
  \BibitemOpen
  \bibfield  {author} {\bibinfo {author} {\bibfnamefont {A.}~\bibnamefont
  {Del~Maestro}}, \bibinfo {author} {\bibfnamefont {B.}~\bibnamefont
  {Rosenow}}, \bibinfo {author} {\bibfnamefont {J.~A.}\ \bibnamefont {Hoyos}},
  \ and\ \bibinfo {author} {\bibfnamefont {T.}~\bibnamefont {Vojta}},\ }\href
  {\doibase 10.1103/PhysRevLett.105.145702} {\bibfield  {journal} {\bibinfo
  {journal} {Phys. Rev. Lett.}\ }\textbf {\bibinfo {volume} {105}},\ \bibinfo
  {pages} {145702} (\bibinfo {year} {2010})}\BibitemShut {NoStop}%
\bibitem [{\citenamefont {Xing}\ \emph {et~al.}(2015)\citenamefont {Xing},
  \citenamefont {Zhang}, \citenamefont {Fu}, \citenamefont {Liu}, \citenamefont
  {Sun}, \citenamefont {Peng}, \citenamefont {Wang}, \citenamefont {Lin},
  \citenamefont {Ma}, \citenamefont {Xue}, \citenamefont {Wang},\ and\
  \citenamefont {Xie}}]{Xingetal15}%
  \BibitemOpen
  \bibfield  {author} {\bibinfo {author} {\bibfnamefont {Y.}~\bibnamefont
  {Xing}}, \bibinfo {author} {\bibfnamefont {H.-M.}\ \bibnamefont {Zhang}},
  \bibinfo {author} {\bibfnamefont {H.-L.}\ \bibnamefont {Fu}}, \bibinfo
  {author} {\bibfnamefont {H.}~\bibnamefont {Liu}}, \bibinfo {author}
  {\bibfnamefont {Y.}~\bibnamefont {Sun}}, \bibinfo {author} {\bibfnamefont
  {J.-P.}\ \bibnamefont {Peng}}, \bibinfo {author} {\bibfnamefont
  {F.}~\bibnamefont {Wang}}, \bibinfo {author} {\bibfnamefont {X.}~\bibnamefont
  {Lin}}, \bibinfo {author} {\bibfnamefont {X.-C.}\ \bibnamefont {Ma}},
  \bibinfo {author} {\bibfnamefont {Q.-K.}\ \bibnamefont {Xue}}, \bibinfo
  {author} {\bibfnamefont {J.}~\bibnamefont {Wang}}, \ and\ \bibinfo {author}
  {\bibfnamefont {X.~C.}\ \bibnamefont {Xie}},\ }\href {\doibase
  10.1126/science.aaa7154} {\bibfield  {journal} {\bibinfo  {journal}
  {Science}\ }\textbf {\bibinfo {volume} {350}},\ \bibinfo {pages} {542}
  (\bibinfo {year} {2015})}\BibitemShut {NoStop}%
\bibitem [{\citenamefont {Ubaid-Kassis}\ \emph {et~al.}(2010)\citenamefont
  {Ubaid-Kassis}, \citenamefont {Vojta},\ and\ \citenamefont
  {Schroeder}}]{UbaidKassisVojtaSchroeder10}%
  \BibitemOpen
  \bibfield  {author} {\bibinfo {author} {\bibfnamefont {S.}~\bibnamefont
  {Ubaid-Kassis}}, \bibinfo {author} {\bibfnamefont {T.}~\bibnamefont {Vojta}},
  \ and\ \bibinfo {author} {\bibfnamefont {A.}~\bibnamefont {Schroeder}},\
  }\href@noop {} {\bibfield  {journal} {\bibinfo  {journal} {Phys. Rev. Lett.}\
  }\textbf {\bibinfo {volume} {104}},\ \bibinfo {pages} {066402} (\bibinfo
  {year} {2010})}\BibitemShut {NoStop}%
\bibitem [{\citenamefont {Vojta}(2010)}]{Vojta10}%
  \BibitemOpen
  \bibfield  {author} {\bibinfo {author} {\bibfnamefont {T.}~\bibnamefont
  {Vojta}},\ }\href {\doibase 10.1007/s10909-010-0205-4} {\bibfield  {journal}
  {\bibinfo  {journal} {J. Low Temp. Phys.}\ }\textbf {\bibinfo {volume}
  {161}},\ \bibinfo {pages} {299} (\bibinfo {year} {2010})}\BibitemShut
  {NoStop}%
\bibitem [{\citenamefont {Vojta}\ and\ \citenamefont
  {Schmalian}(2005{\natexlab{b}})}]{VojtaSchmalian05}%
  \BibitemOpen
  \bibfield  {author} {\bibinfo {author} {\bibfnamefont {T.}~\bibnamefont
  {Vojta}}\ and\ \bibinfo {author} {\bibfnamefont {J.}~\bibnamefont
  {Schmalian}},\ }\href {\doibase 10.1103/PhysRevB.72.045438} {\bibfield
  {journal} {\bibinfo  {journal} {Phys. Rev. B}\ }\textbf {\bibinfo {volume}
  {72}},\ \bibinfo {pages} {045438} (\bibinfo {year}
  {2005}{\natexlab{b}})}\BibitemShut {NoStop}%
\bibitem [{\citenamefont {Cooper}\ \emph {et~al.}(1982)\citenamefont {Cooper},
  \citenamefont {Freedman},\ and\ \citenamefont
  {Preston}}]{CooperFreedmanPreston82}%
  \BibitemOpen
  \bibfield  {author} {\bibinfo {author} {\bibfnamefont {F.}~\bibnamefont
  {Cooper}}, \bibinfo {author} {\bibfnamefont {B.}~\bibnamefont {Freedman}}, \
  and\ \bibinfo {author} {\bibfnamefont {D.}~\bibnamefont {Preston}},\
  }\href@noop {} {\bibfield  {journal} {\bibinfo  {journal} {Nucl. Phys. B}\
  }\textbf {\bibinfo {volume} {210}},\ \bibinfo {pages} {210} (\bibinfo {year}
  {1982})}\BibitemShut {NoStop}%
\bibitem [{\citenamefont {Kim}(1993)}]{Kim93}%
  \BibitemOpen
  \bibfield  {author} {\bibinfo {author} {\bibfnamefont {J.~K.}\ \bibnamefont
  {Kim}},\ }\href@noop {} {\bibfield  {journal} {\bibinfo  {journal} {Phys.
  Rev. Lett.}\ }\textbf {\bibinfo {volume} {70}},\ \bibinfo {pages} {1735}
  (\bibinfo {year} {1993})}\BibitemShut {NoStop}%
\bibitem [{\citenamefont {Caracciolo}\ \emph {et~al.}(2001)\citenamefont
  {Caracciolo}, \citenamefont {Gambassi}, \citenamefont {Gubinelli},\ and\
  \citenamefont {Pelissetto}}]{CGGP01}%
  \BibitemOpen
  \bibfield  {author} {\bibinfo {author} {\bibfnamefont {S.}~\bibnamefont
  {Caracciolo}}, \bibinfo {author} {\bibfnamefont {A.}~\bibnamefont
  {Gambassi}}, \bibinfo {author} {\bibfnamefont {M.}~\bibnamefont {Gubinelli}},
  \ and\ \bibinfo {author} {\bibfnamefont {A.}~\bibnamefont {Pelissetto}},\
  }\href {\doibase 10.1007/BF01352587} {\bibfield  {journal} {\bibinfo
  {journal} {Eur. Phys. J. B}\ }\textbf {\bibinfo {volume} {20}},\ \bibinfo
  {pages} {255} (\bibinfo {year} {2001})}\BibitemShut {NoStop}%
\bibitem [{\citenamefont {Wolff}(1989)}]{Wolff89}%
  \BibitemOpen
  \bibfield  {author} {\bibinfo {author} {\bibfnamefont {U.}~\bibnamefont
  {Wolff}},\ }\href@noop {} {\bibfield  {journal} {\bibinfo  {journal} {Phys.
  Rev. Lett.}\ }\textbf {\bibinfo {volume} {62}},\ \bibinfo {pages} {361}
  (\bibinfo {year} {1989})}\BibitemShut {NoStop}%
\bibitem [{\citenamefont {Metropolis}\ \emph {et~al.}(1953)\citenamefont
  {Metropolis}, \citenamefont {Rosenbluth}, \citenamefont {Rosenbluth},\ and\
  \citenamefont {Teller}}]{MRRT53}%
  \BibitemOpen
  \bibfield  {author} {\bibinfo {author} {\bibfnamefont {N.}~\bibnamefont
  {Metropolis}}, \bibinfo {author} {\bibfnamefont {A.}~\bibnamefont
  {Rosenbluth}}, \bibinfo {author} {\bibfnamefont {M.}~\bibnamefont
  {Rosenbluth}}, \ and\ \bibinfo {author} {\bibfnamefont {A.}~\bibnamefont
  {Teller}},\ }\href@noop {} {\bibfield  {journal} {\bibinfo  {journal} {J.
  Chem. Phys.}\ }\textbf {\bibinfo {volume} {21}},\ \bibinfo {pages} {1087}
  (\bibinfo {year} {1953})}\BibitemShut {NoStop}%
\bibitem [{\citenamefont {Ballesteros}\ \emph
  {et~al.}(1998{\natexlab{a}})\citenamefont {Ballesteros}, \citenamefont
  {Fern\'andez}, \citenamefont {Mart\'in-Mayor}, \citenamefont {Mu\~noz
  Sudupe}, \citenamefont {Parisi},\ and\ \citenamefont
  {Ruiz-Lorenzo}}]{BFMM98}%
  \BibitemOpen
  \bibfield  {author} {\bibinfo {author} {\bibfnamefont {H.~G.}\ \bibnamefont
  {Ballesteros}}, \bibinfo {author} {\bibfnamefont {L.~A.}\ \bibnamefont
  {Fern\'andez}}, \bibinfo {author} {\bibfnamefont {V.}~\bibnamefont
  {Mart\'in-Mayor}}, \bibinfo {author} {\bibfnamefont {A.}~\bibnamefont
  {Mu\~noz Sudupe}}, \bibinfo {author} {\bibfnamefont {G.}~\bibnamefont
  {Parisi}}, \ and\ \bibinfo {author} {\bibfnamefont {J.~J.}\ \bibnamefont
  {Ruiz-Lorenzo}},\ }\href {\doibase 10.1103/PhysRevB.58.2740} {\bibfield
  {journal} {\bibinfo  {journal} {Phys. Rev. B}\ }\textbf {\bibinfo {volume}
  {58}},\ \bibinfo {pages} {2740} (\bibinfo {year}
  {1998}{\natexlab{a}})}\BibitemShut {NoStop}%
\bibitem [{\citenamefont {Ballesteros}\ \emph
  {et~al.}(1998{\natexlab{b}})\citenamefont {Ballesteros}, \citenamefont
  {Fernandez}, \citenamefont {Martin-Mayor}, \citenamefont {Munoz~Sudupe},
  \citenamefont {Parisi},\ and\ \citenamefont {Ruiz-Lorenzo}}]{BFMM98b}%
  \BibitemOpen
  \bibfield  {author} {\bibinfo {author} {\bibfnamefont {H.~G.}\ \bibnamefont
  {Ballesteros}}, \bibinfo {author} {\bibfnamefont {L.~A.}\ \bibnamefont
  {Fernandez}}, \bibinfo {author} {\bibfnamefont {V.}~\bibnamefont
  {Martin-Mayor}}, \bibinfo {author} {\bibfnamefont {A.}~\bibnamefont
  {Munoz~Sudupe}}, \bibinfo {author} {\bibfnamefont {G.}~\bibnamefont
  {Parisi}}, \ and\ \bibinfo {author} {\bibfnamefont {J.~J.}\ \bibnamefont
  {Ruiz-Lorenzo}},\ }\href@noop {} {\bibfield  {journal} {\bibinfo  {journal}
  {Nucl. Phys. B}\ }\textbf {\bibinfo {volume} {512}},\ \bibinfo {pages} {681}
  (\bibinfo {year} {1998}{\natexlab{b}})}\BibitemShut {NoStop}%
\bibitem [{\citenamefont {Zhu}\ \emph {et~al.}(2015)\citenamefont {Zhu},
  \citenamefont {Wan}, \citenamefont {Narayanan}, \citenamefont {Hoyos},\ and\
  \citenamefont {Vojta}}]{ZWNHV15}%
  \BibitemOpen
  \bibfield  {author} {\bibinfo {author} {\bibfnamefont {Q.}~\bibnamefont
  {Zhu}}, \bibinfo {author} {\bibfnamefont {X.}~\bibnamefont {Wan}}, \bibinfo
  {author} {\bibfnamefont {R.}~\bibnamefont {Narayanan}}, \bibinfo {author}
  {\bibfnamefont {J.~A.}\ \bibnamefont {Hoyos}}, \ and\ \bibinfo {author}
  {\bibfnamefont {T.}~\bibnamefont {Vojta}},\ }\href {\doibase
  10.1103/PhysRevB.91.224201} {\bibfield  {journal} {\bibinfo  {journal} {Phys.
  Rev. B}\ }\textbf {\bibinfo {volume} {91}},\ \bibinfo {pages} {224201}
  (\bibinfo {year} {2015})}\BibitemShut {NoStop}%
\bibitem [{\citenamefont {Vojta}\ and\ \citenamefont
  {Hoyos}(2008)}]{VojtaHoyos08b}%
  \BibitemOpen
  \bibfield  {author} {\bibinfo {author} {\bibfnamefont {T.}~\bibnamefont
  {Vojta}}\ and\ \bibinfo {author} {\bibfnamefont {J.~A.}\ \bibnamefont
  {Hoyos}},\ }in\ \href@noop {} {\emph {\bibinfo {booktitle} {Recent Progress
  in Many-Body Theories}}},\ \bibinfo {editor} {edited by\ \bibinfo {editor}
  {\bibfnamefont {J.}~\bibnamefont {Boronat}}, \bibinfo {editor} {\bibfnamefont
  {G.}~\bibnamefont {Astrakharchik}}, \ and\ \bibinfo {editor} {\bibfnamefont
  {F.}~\bibnamefont {Mazzanti}}}\ (\bibinfo  {publisher} {World Scientific},\
  \bibinfo {address} {Singapore},\ \bibinfo {year} {2008})\ p.\ \bibinfo
  {pages} {235}\BibitemShut {NoStop}%
\bibitem [{Note2()}]{Note2}%
  \BibitemOpen
  \bibinfo {note} {For low dilutions $p$, the parabola fits of $g_{\protect \rm
  av}$ vs.\ $L_\tau $ are affected by corrections to scaling for small $L$ and
  $L_\tau $. We thus slightly adjust $L_\tau ^{\protect \rm max}$ and
  $g_{\protect \rm av}^{\protect \rm max}$ to further improve the quality of
  the data collapse onto a common master curve. This applies to the four
  smallest system sizes $L$ for $p=1/8$ and 1/5 and the three smallest sizes
  for $p=2/7$. The resulting change of the value of $z$ is about 0.01, well
  below the error due to the uncertainty in $T_c$.}\BibitemShut {Stop}%
\bibitem [{\citenamefont {Chayes}\ \emph {et~al.}(1986)\citenamefont {Chayes},
  \citenamefont {Chayes}, \citenamefont {Fisher},\ and\ \citenamefont
  {Spencer}}]{CCFS86}%
  \BibitemOpen
  \bibfield  {author} {\bibinfo {author} {\bibfnamefont {J.~T.}\ \bibnamefont
  {Chayes}}, \bibinfo {author} {\bibfnamefont {L.}~\bibnamefont {Chayes}},
  \bibinfo {author} {\bibfnamefont {D.~S.}\ \bibnamefont {Fisher}}, \ and\
  \bibinfo {author} {\bibfnamefont {T.}~\bibnamefont {Spencer}},\ }\href@noop
  {} {\bibfield  {journal} {\bibinfo  {journal} {Phys. Rev. Lett.}\ }\textbf
  {\bibinfo {volume} {57}},\ \bibinfo {pages} {2999} (\bibinfo {year}
  {1986})}\BibitemShut {NoStop}%
\bibitem [{\citenamefont {Prokof'ev}\ and\ \citenamefont
  {Svistunov}(2001)}]{ProkofevSvistunov01}%
  \BibitemOpen
  \bibfield  {author} {\bibinfo {author} {\bibfnamefont {N.}~\bibnamefont
  {Prokof'ev}}\ and\ \bibinfo {author} {\bibfnamefont {B.}~\bibnamefont
  {Svistunov}},\ }\href {\doibase 10.1103/PhysRevLett.87.160601} {\bibfield
  {journal} {\bibinfo  {journal} {Phys. Rev. Lett.}\ }\textbf {\bibinfo
  {volume} {87}},\ \bibinfo {pages} {160601} (\bibinfo {year}
  {2001})}\BibitemShut {NoStop}%
\bibitem [{Note3()}]{Note3}%
  \BibitemOpen
  \bibinfo {note} {We actually use $z=1.45$ which is close to the effective
  dynamical exponent found for the system size range and dilution
  considered.}\BibitemShut {Stop}%
\bibitem [{\citenamefont {Motrunich}\ \emph {et~al.}(2000)\citenamefont
  {Motrunich}, \citenamefont {Mau}, \citenamefont {Huse},\ and\ \citenamefont
  {Fisher}}]{MMHF00}%
  \BibitemOpen
  \bibfield  {author} {\bibinfo {author} {\bibfnamefont {O.}~\bibnamefont
  {Motrunich}}, \bibinfo {author} {\bibfnamefont {S.~C.}\ \bibnamefont {Mau}},
  \bibinfo {author} {\bibfnamefont {D.~A.}\ \bibnamefont {Huse}}, \ and\
  \bibinfo {author} {\bibfnamefont {D.~S.}\ \bibnamefont {Fisher}},\ }\href
  {\doibase 10.1103/PhysRevB.61.1160} {\bibfield  {journal} {\bibinfo
  {journal} {Phys. Rev. B}\ }\textbf {\bibinfo {volume} {61}},\ \bibinfo
  {pages} {1160} (\bibinfo {year} {2000})}\BibitemShut {NoStop}%
\bibitem [{\citenamefont {Altman}\ \emph {et~al.}(2010)\citenamefont {Altman},
  \citenamefont {Kafri}, \citenamefont {Polkovnikov},\ and\ \citenamefont
  {Refael}}]{AKPR10}%
  \BibitemOpen
  \bibfield  {author} {\bibinfo {author} {\bibfnamefont {E.}~\bibnamefont
  {Altman}}, \bibinfo {author} {\bibfnamefont {Y.}~\bibnamefont {Kafri}},
  \bibinfo {author} {\bibfnamefont {A.}~\bibnamefont {Polkovnikov}}, \ and\
  \bibinfo {author} {\bibfnamefont {G.}~\bibnamefont {Refael}},\ }\href
  {\doibase 10.1103/PhysRevB.81.174528} {\bibfield  {journal} {\bibinfo
  {journal} {Phys. Rev. B}\ }\textbf {\bibinfo {volume} {81}},\ \bibinfo
  {pages} {174528} (\bibinfo {year} {2010})}\BibitemShut {NoStop}%
\bibitem [{\citenamefont {Wang}\ \emph {et~al.}(2015)\citenamefont {Wang},
  \citenamefont {Guo},\ and\ \citenamefont {Sandvik}}]{WangGuoSandvik15}%
  \BibitemOpen
  \bibfield  {author} {\bibinfo {author} {\bibfnamefont {Y.}~\bibnamefont
  {Wang}}, \bibinfo {author} {\bibfnamefont {W.}~\bibnamefont {Guo}}, \ and\
  \bibinfo {author} {\bibfnamefont {A.~W.}\ \bibnamefont {Sandvik}},\ }\href
  {\doibase 10.1103/PhysRevLett.114.105303} {\bibfield  {journal} {\bibinfo
  {journal} {Phys. Rev. Lett.}\ }\textbf {\bibinfo {volume} {114}},\ \bibinfo
  {pages} {105303} (\bibinfo {year} {2015})}\BibitemShut {NoStop}%
\bibitem [{\citenamefont {Roscilde}\ and\ \citenamefont
  {Haas}(2007)}]{RoscildeHaas07}%
  \BibitemOpen
  \bibfield  {author} {\bibinfo {author} {\bibfnamefont {T.}~\bibnamefont
  {Roscilde}}\ and\ \bibinfo {author} {\bibfnamefont {S.}~\bibnamefont
  {Haas}},\ }\href {\doibase 10.1103/PhysRevLett.99.047205} {\bibfield
  {journal} {\bibinfo  {journal} {Phys. Rev. Lett.}\ }\textbf {\bibinfo
  {volume} {99}},\ \bibinfo {pages} {047205} (\bibinfo {year}
  {2007})}\BibitemShut {NoStop}%
\end{thebibliography}%

\end{document}